\documentstyle[12pt]{article}
\title{Real space renormalization-group for configurational random walk models
on a hierarchical lattice. The asymptotic end-to-end distance of a weakly
SARW in dimension four.}

\author{
Suemi Rodr\'\i guez-Romo\\Centre of Theoretical Research\\
National University of M\'exico, Campus Cuautitl\'an \\
Apdo. Postal 95, Unidad Militar, Cuautitl\'an Izcalli\\
Estado de M\'exico, 54768 M\'exico $^{\ast}$}

\date{}

\begin{document}

\maketitle

\renewcommand{\thefootnote}{\fnsymbol{footnote}}
\setcounter{footnote}{-1}
\footnote{$\hspace*{-6mm}^{\ast}$
e-mail: suemi$@$servidor.unam.mx \\
\hspace*{1.5cm}suemi@fis.cinvestav.mx}
\renewcommand{\thefootnote}{\arabic{footnote}}\

\baselineskip0.6cm

{\small {\bf Abstract.} We present a real space renormalization-group map
for probabilities of random walks on a hierarchical lattice. From this, we
study the asymptotic behavior of the end-to-end distance of a weakly
self-avoiding random walk (SARW) that  penalizes the (self-)intersection of two
random walks in dimension four on the hierarchical lattice.\

PACS 02.50.+S}

\newpage

\baselineskip0.6cm

\section{Introduction}

Self-avoiding random walk models appeared in chemical physics
as models for long polymer chains. Roughly speaking, a polymer is
composed of a large number of monomers which are linked together
randomly but cannot overlap. This feature is modelled by a self-repulsion
term. Let ${\bf A}$ be a d-dimensional hypercubic lattice, typically
the integer lattice ${\bf Z}^d$ (or a finite subset of ${\bf Z}^d$).
Its elements are called sites, oriented pairs of sites are called steps.
A walk $w$ on ${\bf A}$ is an ordered sequence $(w(0),w(1),...,w(k))$ of
sites in  ${\bf A}$; $k\geq 0$, $k$ being the length of the walk \cite{Fer}.
Thus, a simple random walk on ${\bf A}$, starting at $w(0)\in {\bf Z}^d$,
is a stochastic process indexed by the non-negative integers.\

To state a self-repulsion term, Flory \cite{Fr} used the self-avoiding walk
(SAW). A self-avoiding walk of length $n$ is a simple random walk which
visits no site more than once. Although this simple model possesses some
qualitative features of polymers, turns out to be very difficult for obtaining
rigorous results. Instead of using SAW we take advantage of the
several measures on random walks which favor self-avoiding walks.\

In this paper we study weakly SARW (so called Domb-Joyce model or
self-repellent walk). This is a measure on the set of simple walks in which
self-intersections are discouraged but not forbidden. Here, double
intersections
of random walks are penalized by a factor
$e^{-\lambda \mbox{ (measure of (self-)intersection)}}$, $\lambda>0$ being a
small constant.
This factor is needed to make the process tend to avoid itself. Thus,
probabilities of the random walks are modified by a measure of
(self-)intersection inside the lattice. This measure is written in terms of
waiting (or local) times for the process ($t_i$). Here, as was done in
reference
[5], the measure is $\sum_{i<j}t_it_j$, for all local times in the\newline
(self-)intersections of the walks on the lattice, which can be either
hypercubic
or hierarchical. If in this model the lattice is the hypercubic lattice, then
the state space is ${\bf Z}^d$. To understand this model, we use a
hierarchical lattice which state space is defined in next section. We want to
stress that the only feature of the model that depends upon the lattice used
(hypercubic or hierarchical) is the state space in this. Thus, definitions
for the weakly SARW in section 4 (interaction energy in Theorem 4) are
equivalent in both cases, but with different state space in the lattice. We
develop our method on a hierarchical lattice because they have the feature
that the renormalization-group map is particulary simple, which is not the
case on the hypercubic lattice. We believe that the results of this procedure
extend to weakly SARW on the hypercubic lattice.\

The hierarchical models introduced by Dyson \cite{Dy} feature a simple \newline
renormalization-group transformation. This can easily be seen  in related
literature $\cite{Ga}$. We would like to understand the logarithmic correction
that appears in the hypercubic lattice, d=4, for the end-to-end distance
of the weakly SARW. The real space renormalization-group map we develop
here, is factorizable only in terms of a hierarchical lattice. So, we present
a method in which $A$ is labeled in terms of a hierarchical metric space,
from this an easy realization of the map is followed. In the integer lattice
${\bf Z}^d$, this is not true and technical problems arise.\

We use a hierarchical lattice where the points are labeled by elements of a
countable, abelian group ${\it G}$ with ultrametric $\delta$; i.e. the metric
space $({\it G},\delta)$ is hierarchical. The hierarchical structure of this
metric space induces a renormalization-group map that is ``local"; i.e.
instead of studying the space of random functions on the whole lattice, we
can descend to the study of random functions on L-blocks (cosets of {\it G})
\cite{Ev}. This simplifying feature was used by Brydges, Evans and Imbrie
\cite{Ev} to prove (in the $\lambda \phi^4$ Grassmann valued field
representation for a weakly SARW that penalizes the (self-)intersection of
two random walks) that the introduction of a sufficiently weakly
self-avoidance interaction does not change the decay of the Green's function
for a particular L\'{e}vy process (continuous time random walk), \cite{Ev}
when d=4, provided the mass is introduced critically.\

A rigorous proof of the end-to-end distance for the weakly SARW, $d=4$, on
the hierarchical lattice, has recently been reported \cite{Im}. This was done
in
the field theoretical approach by means of controlling the interacting
Green's function and inverting the Laplace transform.\

Low dimensional models are the most interesting from physical viewpoint,
but rigorous results are difficult of being obtained. One major result is the
proof that in high dimensions the exponents of weakly SARW take the
``mean-field" value \cite{Ma}. In this context ``lace expansion"  is one of
the most successful tools. Contrary to what has been done from this method,
in this paper we develop a map on real space which is full of the physical
intuition
needed for being applied on some other cases not yet solved.\

Weakly SARW exhibits logarithmic corrections for physically meaningful
magnitudes in the critical dimension of the model, i.e. d=4. We study the
end-to-end distance for weakly SARW that penalizes the double crossing of
walks in $d=4$. A probabilistic meaning is given to the exponent of this
logarithmic correction. In this paper we present an heuristic space-time
renormalization-group argument to show that the end-to-end distance of a
weakly self-avoiding random walk (SARW) on the hierarchical lattice, that
penalizes the (self-)intersection of two walks in $d=4$, is asymptotic to a
constant times $T^{\frac{1}{2}}log^{\frac{1}{8}}T$ as T tends to  infinity,
T being the total time for the walk. This has already been conjectured
before \cite{Br}. This is the testing ground to check that our map reproduces
previously known results with physical improved intuition. The weakly
self-avoiding random walk model that penalizes the (self-)intersection of two
random
walks is in the same universality class as the perfect  self-avoiding random
walk model; therefore, the same logarithmic correction for the end-to-end
distance is expected to hold for both cases regardless the details of the
state space used. Real space renormalization-group methods have proved to be
useful in the study of a wide class of phenomena.\

Since our method is intended to provide an alternative way, full of physical
intuition, to renormalize random walk models on a hierarchical lattice; we
study weakly SARW in $d=4$ just as a testing ground. We consider our method
suitable of being directly applied on kinetically growing measure models,
discrete version. Kinetically growing measure models are produced from
consistent measures and are the natural framework for random walks provided
these are seen like stochastic processes. Myopic self-avoiding walk (a model
for adsorption of linear polymers on surfaces \cite{Bu}), infinite
self-avoiding walk and Laplacian (or loop-erased) random walk \cite{La} are
examples of kinetically
growing measure models. These models have been studied mainly in the field
theoretical framework where has been shown \cite{De} that some contributions
neglected in the derivation of the continuum problem from the discrete version
of the model might be essential in determining the asymptotic behaviour of the
model analytically. In fact, usually the corresponding action (in the
continuum limit) is not fully renormalizable. We address this problem by
presenting a real space renormalization-group suitable of been applied on
discrete models. The metod can be used for obtaining both, rigorous and
heuristic results for these models.\

We hope this paper could be interesting for various readerships. We intend to
brydge a gap between probabilistic approaches and field theoretical ones,
thereby providing a new probabilistic meaning to critical and asymptotic
exponents. The method we develop in this paper can be used also as an
intuitive mean to search new exact results, which then remain to be proven
rigorously within the framework presented here, or by other means. The
renormalization scheme constructed here is not considered or known in the
literature.\

This paper is organized as follows; in Section 2 we present
the hierarchical lattice and the L\'{e}vy process we study, on the space
of simple random walks. In Section 3 we define the
renormalization-group map on the hierarchical lattice and prove that two
particular probabilities, functions of random walks, flow to fixed forms
after applying the map. In Section 4 we apply the renormalization-group map
to the weakly SARW model that penalizes the intersection of two random walks
on the hierarchical lattice. In Section 5 we present an heuristic
proof for the asymptotic behavior of the end-to-end distance for the weakly
SARW on the hierarchical lattice, d=4. Although this is an heuristic proof,
it helps in understanding the way the map can be used and gives a new
probabilistic meaning to the exponent of the logarithmic correction. Sections
3, 4 and 5 involve important results of this paper, being Theorem 4
the main result reported in the paper and an important step to obtain
heuristically the end-to-end distance for the weakly SARW. Even more, we claim
this Theorem to be a random walk version of the field theoretical
approach in reference [5] with improved physical intuition. Finally, we
summarize.\

\section{The hierarchical random walk.}

The hierarchical lattice used in this paper was recently introduced by
Brydges, Evans and Imbrie \cite{Ev} \cite{Im}. Here we are presenting a slight
variant of the model in reference [5].\

Fix an integer $L\geq 2$. Hereafter, the points of the lattice ${\bf A}$ are
labeled by elements of the countable abelian group
${\it G}=\oplus ^{\infty}_{k=0}{\bf Z}_{L^d}$, d being the dimension
of the lattice. Through the paper the abelian group
${\it G}=\oplus ^{\infty}_{k=0}{\bf Z}_{L^d}$ replaces ${\bf A}$.\

 An element $X$ in ${\it G}$
is an infinite sequence
$$
X\equiv (...,X_k,...,X_2, X_1, X_0 )\;\; ;
\mbox{$X_i \in {\bf Z}_{L^d}$ thus $X\in {\it G}=\oplus^{\infty}_{k=0}
{\bf Z}_{L^d}$},
$$
where only finitely many $X_i$ are non-zero.\

Let us define subgroups
\begin{equation}
\{0\}={\it G}_0\subset {\it G}_1\subset ...\subset {\it G}
\;\;\;\mbox{where  }
{\it G}_k=\{X\in {\it G}| X_i=0, i\geq k \},
\end{equation}
and the norm $|\cdot|$ as
\begin{equation}
|X|=\left\{\begin{array}{cc}
0 & \mbox{ if $X=0$} \\
L^p & \mbox{where   $p=\inf\{k| X\in {\it G}_k\}$  if   $X\neq 0$}.
\end{array}
\right.
\end{equation}
Then, the map $\delta:(X, Y)\rightarrow |X-Y|$ defines a metric on ${\it G}$.
In this metric the subgroups
${\it G}_k$ are balls $|X|\leq L^k$ containing $L^{dk}$ points.
Here the operation + (hence - as well) is defined componentwise.\

In Figure 1 we have described two examples for $L=2$, a one-dimensional
hierarchical lattice (Figure 1.a)) already presented in reference [5] and a
two-dimensional hierarchical lattice (Figure 1.b)). In these Figures we
depict ${\it G}_k$ cosets, a way to calculate distances among
points, and the concept of scales for each example.\

The metric defined by eq(2) satisfies a stronger condition than the
 triangle inequality, namely
\begin{equation}
|X+Y|\leq \mbox{ Max}(|X|,|Y|).
\end{equation}
This {\it non-archimedean} property implies that every triangle is isosceles
and that every point interior to a ball can be considered its center.
Moreover, balls of radius L are the same as balls of diameter L, and are
the same as ${\it G}_1$ cosets. From inequality (3), it is clear that  the
metric introduced is an ultrametric  and confers the hierarchical structure.
Strictly speaking, it is only the metric space $({\it G},\delta)$ that is
hierarchical. Here, ultrametric appears naturaly as a property of polynomials.
It can be shown that ${\it G}_k$ represents polynomials of degree $k$ on a
formal basis.\

Let us now introduce the L\'{e}vy process we propose in this paper.
The elements of the lattice ${\it G}$ are called sites; unoriented pairs
$\{X,Y\}$ of sites in ${\it G}$ with $X\neq Y$ are called bonds; oriented
pairs $(X,Y)$ are called steps (or jumps) with initial site $X$ and final
site $Y$. Let us define the L\'{e}vy process \cite{Ev}($\equiv $ continuous
time random walk), $w$, as an ordered sequence of sites in {\it G};
\begin{equation}
(w(t_0),...,w(t_0+...+t_n))\;\;,w(t_0+...+t_i)=X_i\in {\it G},
\;\;T=\sum ^n_{i=0}t_i ,\; n\geq 0
\end{equation}
where $t_i$ is the time spent in $X_i\in {\it G}$ (waiting time at $X_i$)
and T, fixed, is the running time for the process. For convenience we take
$X_0=0$. The support of the walk $w$ is defined by
\begin{equation}
supp(w)=\{X\in {\it G} | w(t_0,...,t_j)=X \mbox{ for some j}\},
\end{equation}
for any $w$. The random walk we are dealing with is not the nearest
neighbor random walk on the lattice, provided we mean neighbourhood with
respect to the ultrametric distance $\delta$ previously defined.\

If we compare this L\'evy process with the simple random walk in the
hypercubic lattice defined in the Introduction, we see that in this Section we
construct a
different stochastic process. Here, time is continuous and the path of the
walk is given on ${\it G}$ (in the hypercubic lattice, the walk was indexed
by non-negative integers, instead of waiting times, with path on ${\bf Z}^d$).
Besides, here we fix the running time for the process. This corresponds to
walks of fixed length, feature that was not imposed on simple random walks on
the hypercubic lattice.\

We suppose the L\'{e}vy process in the hierarchical lattice has a probability
$rdt$ (r is the jumping rate) of making a step in time $(t, dt)$ and, given
that it jumps, the probability of jumping from X to Y is $q(X,Y)$. Thus, the
process, conditioned to $n$ jumps spaced by times $t_0,t_1,...,t_n$, has a
probability density
\begin{equation}
P(w)=r^ne^{-rT}\prod^{n-1}_{i=0}q(X_{i+1},X_{i})
\;\;\mbox{, where $T=\sum^{n}_{i=0}t_i$.}
\end{equation}
Define $Dw$ by
$$
\int(\cdot ) Dw=
\sum_{n}\;\;\sum_{\left[ X_{i}\right]^{n}_{i=0}}
\;\;\;\int^{T}_{t_i=0}\;\;\prod^{n}_{i=0}\;\;dt_{i}
\delta(\sum^{n}_{i=0}t_{i}-T)(\cdot ).
$$
{}From this and eq(6) it is straightforward to obtain
$$
\int P(w)Dw=
\left\langle \sum_{\left[ X_{i}\right]^{n-1}_{i=0}}
\prod^{n-1}_{i=0}q(X_{i+1},X_{i})
\right\rangle_{Poisson}=1,
$$
$\left\langle (\cdot) \right\rangle$ being the expectation of ($\cdot$).\

Here $\prod^{n-1}_{i=0}q(X_{i+1},X_{i})$ has been normalized on
${\it G}$ and we have used
$$
\int^T_0\prod^n_{i=0}r^ne^{-rt_i}\prod^n_{j=0}dt_j\delta
\left(\sum^n_{j=0}t_j-T\right)=
$$
$$
r^ne^{-rT}\int^T_0\prod^n_{i=0}dt_i\delta
\left(\sum^n_{i=0}t_i-T\right)=\frac{\left(rt^n\right)}{n!}e^{-rT}.
$$
\section{The renormalization-group map.}

To start with, let us introduce a renormalization-group map on the
lattice; $R(X_i)=LX'_i$ where $X_i\in {\it G}$ and $LX'_{i}\in {\it G}'$
=${\it G}/{\it G}_1\sim {\it G}$; i.e.
the renormalized lattice ${\it G}'$ is isomorphic to the original
lattice ${\it G}$.\

{}From this renormalization-group map we construct $R(w)=w'$, from $w$
above as defined, to $w'$. Here,
$w'$ is the following ordered sequence of sites in ${\it G}'=
{\it G}/{\it G}_1\approx {\it G}$;
\begin{equation}
(w'(t'_0),...,w'(t'_0+...+t'_k))\;\;,\mbox{ where}
\end{equation}
$$
w'(t'_0,...,t'_{i'})=X'_{i'}\in {\it G},\;\;T'=\sum ^k_{i'=0}t'_i ,
\;0\leq k\leq n,
\;\;T=L^{\beta}T'.
$$
R maps $w(t_0),w(t_0+t_1),...,w(t_0+...+t_n)$ to cosets
$w(t_0)+{\it G}_1,w(t_0+t_1)+{\it G}_1,...,w(t_0+...+t_n)+{\it G}_1$. If
two or more successive cosets in the image are the same, they are listed only
as one site in $w'(t'_0),...,w(t'_0+...+t'_k)$, and the times $t'_j$ are sums
of the corresponding $t_j$ for which successive cosets are the same, rescaled
by $L^{\beta}$. For $\beta=2$, we are dealing with normal diffusion,
in case $\beta<2$ with superdiffusion, and subdiffusion for $\beta>2$.
Additionally, the renormalization-group maps each ${\it G}_1$ coset to
the center of the ball and rescales by $L$. In reference [5], $\beta$ is set
to 2.\

The renormalization group map, applied to functions of the hierarchical
random walk, preserves locality \cite{Ev}. Thus, if
$F(w)=\prod_{X_{i}\in {\it G}}f(w(t)=X_{i})$, the effect of the
renormalization-group
map on $F(w)$ can be studied as the product,
for all elements of the group ${\it G}/{\it G}_1={\it G}'\sim {\it G}$, of
the images of the renormalization-group map of $f(w)$ in the ${\it G}_1$
coset \cite{Ev}. This can be seen as follows; $\prod_{X_{i}\in {\it G}}$
splits into two parts; the first part is $\prod_{X'_{i'}\in {\it G}/{\it G}_1}$
which corresponds to $\prod_{X'_{i'}\in {\it G}}$. The second part is
$\prod_{X_{i}\in ({\it G})_{X'_{i'}}}$ and stands for the $L$-block (${\it
G}_1$
coset) of the lattice which, under the renormalization-group transformation,
maps to $LX'_{i'}$, $(0\leq i'\leq k)$. The geometrical interpretation of this
is quite simple. The renormalization-group map applied on $F(w)$ in ${\it G}$
splits into that of $f(w(t)=X_{i})$ in the contracting ${\it G}_1$ cosets,
multiplied by the whole ${\it G}/{\it G}_1$ group. To study the
renormalization of $F(w)$ in ${\it G}$ we descend to analize the
renormalization-group action on $f(w(t)=X_{i})$ in the contracting ${\it G}_1$
coset.\

In Figure 2 we present the lattice ${\it G}$. On this, walks with fixed
topology in ${\it G}/{\it G_1}$ are depicted. In this example three different
types of local topologies inside the ${\it G}_1$ cosets are illustrated.
Once the renormalization-group map is applied, the renormalized lattice and
random walk are also shown. The renormalized random walk $w'$ visits each
type of ${\it G}_1$ cosets once, twice and three times respectively. In the
renormalized lattice we show the contracted ${\it G}_1$ cosets. The particular
fixed topology chosen for $w'$ is one in the class of the simplest cases
studied by our method. Here is clear what is meant by factorization and
locality of the renormalization-group map. The map factorizes into two terms.
Roughly speaking, the first term corresponds to events inside $L$-blocks or
${\it G}_1$ cosets and the second term corresponds to events outside the
$L$-block (${\it G}_1$ cosets); therefore in ${\it G}/{\it G}_1$=$
{\it G}'\sim {\it G}$. Moreover, for obtaining the flow of the interaction
constant, the map descends to study events in ${\it G}_1$ cosets; thereby
preserving locality.\

We can now work out probabilities at the $(p+1)^{th}$ stage in the
renormalization provided only that we know the probabilities at the $p^{th}$
stage. We sum over the probabilities of all the walks $w^{(p)}$ consistent
with a fixed walk $w^{(p+1)}$ in accordance with the following;\

{\bf Definition. Let $R(w)=w'$ be the renormalization-group
map, above as stated. Then
\begin{equation}
P'(w')=
L^{\beta k}\int Dw P(w)\chi (R(w)=w')
\end{equation}
for any probability  $P(w)$ where the running time for the process
$T=\sum^{n}_{i=0}t_{i}$ is fixed}.\

In this definition, R is a renormalization-group transformation that maps
a density $P(w)$ to a new one, $P'(w')$, on rescaled coarse walks.
Besides, $\chi(c)$ is the characteristic function of the
condition c.\

Let $P(w)=\prod_{X\in {\it G}}p(w(t)=X)$ then\newline
$P'(w')$=$L^{\beta k}\int Dw\prod_{X\in {\it G}}p(w(t)=X)\chi (R(w)=w')$ .
{}From this and factorization properties in the hierarchical lattice follows
$$
P'(w')=\prod_{X'\in{\it G}}\left\{
L^{\beta k}\int Dw\prod_{X\in ({\it G}_1)_{X'}}p(w(t)=X)\chi(R(w)=w')\right\}
$$
$$
=
\prod_{X'\in {\it G}}p'(w'(t')=X'),
$$
that proves in this case the statement of preservation of locality as above
given for $F(w)$. Eq(17) and eq(21) are examples for suitable $P(w)$.\

Eq(8) corresponds precisely to
\begin{equation}
P'(w')=
L^{\beta k}\;\sum_{\left[ n_{i'}\right]^{k}_{i'=0}}
\;\sum_{\left[ X_{i}\right]^{n}_{i=0}}
\int \prod^{n}_{i=0}dt_{i}\;
\prod^{k}_{j'=0}\;
\delta (\sum^{m_{j'}}_{{i}=m_{j'-1}+1}\;\;t_{i}-
L^{\beta }t'_{j'})\times
\end{equation}
$$
\times
\prod^{k}_{j'=0}\;
\prod^{m_{j'}}_{{i}=m_{j'-1}+1}\;
\chi (X_{i}\in LX'_{j'}\;\;)P(w).
$$
Hereafter
\begin{equation}
m_{j'}=\sum^{j'}_{i'=0}n_{i'}+j' \;\;\mbox{ and}
\end{equation}
$$
n=\sum^{k}_{i'=0}n_{i'}+k\;\;\;\;\;\;0\leq j'\leq k.
$$
$n_{i'}$ is the number of steps (for walks $w$) in the ${\it G}_1$
coset which, once the renormalization-group map is applied, has the image
$LX'_{i'}$. Thus, the total number of steps (for walks $w$) on the lattice is
given by the sum of the steps within each L-block (${\it G}_1$ cosets) plus
$k$ times 1 (due to the step out of the corresponding block).\

{\bf Theorem 1. The probability  in eq(6), where $q(X_{i+1},X_{i})$ is
chosen of the form $c|X_{i+1}-X_{i}|^{-\alpha}$ (c is a constant fixed up to
normalization and $\alpha$ another constant), is a fixed point of the
renormalization-group map $R(w)=w'$, provided $\beta =\alpha-d$}.\

{\bf Proof}. Using the definition of the renormalization-group
map  on the probability given in eq(6) and doing some elementary
manipulations we arrive to the following expression
\begin{equation}
P'(w')=L^{\beta k}\;r^{k}\;e^{-rT}
\prod^{k-1}_{j'=0}\;\;(q(LX'_{j'}-
LX'_{j'+1})L^d)\times
\end{equation}
$$
\prod^{k}_{j'=0}\;\;\sum_{n_{j'}}\;\;
r^{n_{j'}}\;\;(q_{1}(L^d-1))^{n_{j'}}
\frac {(L^{\beta }t'_{j'})^{n_{j'}}}
{n_{j'}!}
$$
where $\prod^{n-1}_{i=0}q(X_{i+1}-X_{i})$ has been split into two factors; the
first factor
corresponding to jumps from one $L$-block to another $L$-block
(different ${\it G}_1$ cosets) and the second factor corresponding
to jumps inside the same $L$-block or ${\it G}_1$ coset. Function $q_1$ in
eq(11)
is the probability of jumping to a given point within the ${\it G}_1$ coset
that has the image $LX'_{j'}$
(hereafter $\left({\it G}_{1}\right)_{X'_{j'}}$).
 There are $(L^d-1)$ possibilities with equal probability $q_1$ and
$n_{j'}$ steps.\

{}From normalization, i.e. $\sum_{X\in {\it G}}q(X)=1$,
we get $c=\frac {L^{\alpha -d}-1}{1-L^{-d}}$ and \newline
$q_{1}(L^d-1)=1-L^{d-\alpha }$.\

On the other hand, we know that $T=L^{\beta }T'$, therefore, eq (11) becomes
\begin{equation}
P'(w')=L^{(d+\beta -\alpha)k}\;\;r^{k}
\;\;\prod^{k-1}_{i'=0}q(X'_{i'+1}-X'_{i'})e^{-rL^{(d+\beta -\alpha)}T'}
\end{equation}
Provided $d+\beta -\alpha =0$, eq(12) is clearly a fixed point
of the renormalization-group map $R$, i.e.
$P'(w')=r^{k}\prod^{k-1}_{i'=0}q(X'_{i'+1}-X'_{i'})e^{-rT'}$.

					      Q. E. D.\

Theorem 1 corresponds to the case worked out by Brydges, Evans,
and Imbrie if we choose $\beta =2$, i.e. diffusive
behavior \cite{Ev}.\

{\bf Theorem 2. If the probability  in eq(6) where
$q(X_{j},X_{j+1})$ is chosen of the form
\begin{equation}
q(X_j-X_{j+1})=c\left(\frac{1}{|X_j-X_{j+1}|^\alpha }+
\frac{1}{|X_j-X_{j+1}|^\gamma }\right)
\end{equation}
(c is a constant fixed up to normalization, $\alpha$ and $\gamma$ are
constants, $\alpha\neq \gamma$) then $P(w)$ flows to the fixed
point $P'(w')$ given in theorem 1, (i.e. $q(X_j-X_{j+1})$
is given as in theorem 1), under the
renormalization-group map $R(w)=w'$,
provided $\gamma >>\alpha$ (such that
$log \left(\frac{L^{-\alpha}-L^{-\gamma}-2L^{d-\gamma-\alpha}}{L^{-\alpha}-
2L^{d-\gamma}}\right)\rightarrow 0$) and $\beta =d+\alpha $}.

{\bf Proof}. Following the same ideas as in theorem 1, from normalization,
we obtain
$
c=\frac{(1-L^{d-\alpha})(1-L^{d-\gamma})}
{(L^d-1)(L^{-\alpha}+L^{-\gamma}-2L^{d-\gamma -\alpha})}
\;\;\mbox{  and  }
q_{1}=c(L^{-\gamma}+L^{-\alpha}) .
$
Then, if $\gamma >> \alpha$, $P'(w')$
corresponds to
\begin{equation}
P'(w')=L^{(d+\beta -\alpha)k}\;\;r^{k}
\;\;e^{-rL^{\beta }T'}\times
\end{equation}
$$
\times
\prod^{k-1}_{j'=0}\frac
{(L^{\alpha -d}-1)}{(1-L^{-d})}
|X'_{j'}-X'_{j'+1}|^{-\alpha}
e^{r(1-L^{d-\alpha})L^{\beta }T'}
$$
If $d+\beta -\alpha=0$, eq(14) reduces to $P'(w')$ as given in
Theorem 1.\

					    Q.E.D.\

Theorem 2 is presented here in order to learn about the L\'{e}vy process we
are studying. This case is intended to answer the question about the
feasibility of introducing perturbations in the probability and the possible
results to be obtained. We are looking forward to studying of asymmetric and
``trapping" environments .\

\section{The renormalization-group map on \newline
the weakly SARW}

A  configurational random walk model can be defined by assigning to every
n-tuple of walks
$w_1,...,w_n $ $ (n\geq 0)$ a statistical weight. For a simple random walk
model, this is the product of the statistical weights for each of the n walks
and can serve as a random walk representation of the Gaussian model
\cite{Fer}\cite{Br}. The best known mathematical model that involves a
self-repulsion term is the self-avoiding random walk. A self-avoiding
walk of length n is a simple random walk which visits no site more
than once.
Unfortunately, it turns out that it is extremely difficult to obtain rigorous
results from this model for $d\leq 4$ \cite{Ma}\cite{La1}
\cite{La2}. However, there are other
ways  to include self-repulsion terms in random walk models \cite{La}.
 These split  naturally into two categories: configurational measures where
random walks are weighted  by the number of (self-)intersections, and
kinetically growing measures where random walks are produced from
consistent measures that are generated by Markovian transition probabilities
on the states space of simple random walks (these measures are non-Markovian
on the state space of the lattice) \cite{La}. The weakly self-avoiding random
(or Domb-Joyce) model and the Edwards model are examples of the first category
\cite{La}. The ``true" (or ``myopic") self-avoiding walk and the Laplacian
random walk are examples of the second category \cite{La}.
In this paper we deal only with configurational measures.

A simple random walk model can be thought of as being endowed with a
configurational measure where the weight for the self-intersections of a
walk (and/or among the n-tuple of walks $w_1,...,w_n $ $ (n\geq 0)$) is null.
Configurational measures are measures on $\Lambda_n$, the space of simple
random walks of length n. Let $P_U(w)$ be a probability on $\Lambda_n$
such that
\begin{equation}
P_U(w)=\frac{U(w)P(w)}{Z}
\end{equation}
where $Z=\int U(w)P(w)Dw$. In this paper we consider $P(w)$ as the
probability on $\Lambda_n$ given in eq(6) and $U(w)$ as the interaction
energy of the walks \cite{Fer}. Thus, to study the effect of the
renormalization-group map on $P_U(w)$ we need to follow the trajectory of
$U(w)$ after applying several times the renormalization-group map.\

Therefore, from the definition of the renormalization-group
map in eq(8);
$$
P'_{U'}(w')=L^{\beta k}\int
P_{U}(w)
\chi (R(w)=w')Dw
$$
where $Z'=Z$, it follows that
\begin{equation}
U'(w')=
\frac{\int Dw P(w)\chi (R(w)=w')U(w)}
{\int Dw P(w)\chi (R(w)=w')}
\end{equation}
Note that eq(16) can be viewed as the expectation of $U(w)$
given that the renormalization-group map is imposed, calculated using
$P(w)$ on $\Lambda_{n}$ defined as in eq(6). Therefore, and hereafter,
to simplify notation, we write eq(16) as $U'(w')=< U(w) >_{w'}$.\

 In this Section we deal with $U(w)$ factorizable in terms of the interaction
energy with null weight
for the (self-)intersection of $n$ $(n=2,3,...,etc.)$ random walks
(i.e. a simple random walk factor), and the
interaction energies that
weight the intersection of n random walks. Hereafter, as a hypothesis,
we assume all the factors in
$U(w)$ (functions of $w$) as independent, simple, random variables.
Thus, the conditional (given the $R(w)=w'$ map) expectation of
$U(w)$ is the product of conditional (given the $R(w)=w'$ map)
expectation of each factor in $U(w)$.
This hypothesis follows the same spirit as in the approach used
in polymer networks \cite{Du} and can be seen as a consequence of
factorizability and locality of the map. See Figure 2.\

To start with, we study the simple random walk model such that
\begin{equation}
U(w)=\prod_{X\in {\it G}} e^{-a\sum _{j\in J_{X}}\;\;t_j},
\end{equation}
where $J_{X}=\{j\in \{0,...,n\}|X_j=X \}$ for $w(t_0+,...,+t_j)=X_j$ and
$X\in {\it G}$.\

{\bf Theorem 3. The probability  $P_U(w)$ for the simple random walk
model where
U(w) is given by eq(17), is a fixed form of the \newline
renormalization-group map
$R(w)=w'$ such that, after applying the renormalization-group map,
$a'=L^{\beta }a$}.\

{\bf Proof}. Let us split the product on sites in the lattice in
$ \prod_{X\in {\it G}} e^{-a\sum _{j\in J_{X}}\;\;t_j} $
into two parts. The first one, i.e.
$
\prod_{X'_{i'}\in {\it G}/{\it G_1}}
\mbox{ corresponds to }
\prod_{X'_{i'}\in {\it G}}
$
due to the hierarchical structure of the lattice. The second one, i.e.
$
\prod_{X_{i}\in ({\it G_1})_{X'_{i'}}}
$
stands for the L-block (${\it G_1}$ coset) of the lattice ${\it G}$
that, under the renormalization-group transformation, maps to
$LX'_{i'},\;\;(0\leq i'\leq k)$. There
are $k$ replicas of this. If we again split $\prod^{n-1}_{i=0}q(X_{i+1}-X_i)$
into two factors,
as was done in theorem 1. We obtain
\begin{equation}
U'(w')=\prod_{X'_{i'}\in {\it G}}
\left\{
\prod_{j'\in J_{X'_{i'}}}
e^{q_{1}(L^d-1)r(L^{\beta}t'_{j'})}
\right\}^{-1}\times
\end{equation}
$$
\left\{
\sum_{n_{i'}}\;\int\prod_{i\in I_{X'_{i'}}}
\;dt_{i}\;\prod_{j'\in J_{X'_{i'}}}
\;\delta (\sum^{m_{j'}}_{{i}=m_{j'-1}+1}\;\;t_{i}-
L^{\beta }t'_{j'})\times
\right.
$$
$$
\sum_{X_{i}\in ({\it G}_1)_{X'_{i'}}}
\prod^{m_{j'}}_{{i}=m_{j'-1}+1}\;\;\;
\chi (X_{i}\in LX'_{j'})\;\;
(q_{1}(L^d-1)r)^{n_{i'}}
\left.
e^{-a\sum_{\left(X_{i}\in {\it G_1}\right)_{X'_{i'}}}
\sum _{j\in J_{X_{i}}}\;\;
t_{j}}
\right\}
$$
where we have defined , for $X_{i}\in w$
and $X'_{i'}\in w'$;
\begin{equation}
I_{X'_{i'}}=\left\{i\;|\;X_{i}\in
LX'_{i'}\right\}=
\end{equation}
$$
\bigcup_{j'\in J_{X'_{i'}}}
\left\{i\;|\;m_{j'-1}+1\leq i\leq m_{j'}\right\}=
\bigcup_{j'\in J_{X'_{i'}}}
\left\{i\;|\; 0\leq i\leq n_{j'}\right\}.
$$
Rearranging the double sum in the exponential of eq(18), we obtain
\begin{equation}
U'(w')=\prod_{X'\in {\it G}'}
e^{-aL^\beta \sum_{j'\in J_{X'}}t'_{j'}}.
\end{equation}
Q.E.D.\

The next model we want to study is a weakly self-avoiding random walk (or
Domb-Joyce model), with a configurational measure in which double
intersections of walks are penalized by (roughly speaking) a factor of
$e^{-\lambda}$. As $\lambda \rightarrow \infty$, this reduces to random walks
with strict mutual avoidance. Recall that this weakly model (with
$\lambda >0$) and the perfect self-avoiding random walk, are in the same
universality  class (this implies that the critical exponents are the same).
If $\lambda=0$, this corresponds to a simple random walk model. Next Theorem
is an important result of this paper; it is a random walk version of the
field theoretical approach \cite{Ev} with improved physical intuition and the
key stone in our method to obtain the end-to-end distance of the weakly SARW.
The renormalization-group map applied on a weakly SARW involves the
paramaters $\gamma_1$, $\gamma_2$, $\beta_1$, $\eta$, $A$, $B$, $C$. These
are some conditional expectations of local times for different topologies in
both, $w$ and $w'$ random walks and are precisely defined in Figure 3.\

In Figure 3, $\gamma_1$ and $\gamma_2$ correspond to $O(\lambda)$ and
$O(\lambda^2)$ contributions  in Figure 2.b), respectively. $\beta_1$
corresponds to $O(\lambda^2)$ contribution in figure 2.c) and $\eta$
corresponds to $O(\lambda^2)$ contribution in Figure 2.d). Although $A$, $B$,
and $C$
are not depicted in Figure 2, it is straightforward to figure out the
corresponding pictures.\

{\bf Theorem 4. For the weakly SARW with interaction energy}
\begin{equation}
U(w)=\prod_{X_i\in {\it G}}
e^{-\xi\sum_{i\in J_{X_{i}}}t_{i}-\lambda \sum_{i<j\in J_{X_{i}}}
t_{i}t_{j}
{\bf 1}_{\left\{w(t_i)=w(t_j)\right\} }}
\end{equation}
{\bf $\xi_{2}<0$ and $\lambda >0$ being (small) constants, the probability
 $P_U(w)$ flows to a fixed form after the renormalization-group
 transformation is applied. This fixed form
is characterized by the interaction energy}
\begin{equation}
U'(w')=
\prod_{X'_{i'}\in{\it G}}
e^{
-\xi'
\sum_{i_{\alpha_{1}}'\in J_{X'_{i'}}}
t'_{i_{\alpha_{1}}'}-
\lambda'
\sum_
{\stackrel{i_{\alpha_{1}}'< j_{\alpha_{1}}'}
{\left\{i_{\alpha_{1}}',j_{\alpha_{1}}'\right\}\in J_{X'_{i'}}}}
t'_{i_{\alpha_{1}}'}t'_{j_{\alpha_{1}}'}
{\bf 1}_{(w(t'_{i'_{\alpha_{1}}})=w(t'_{j'_{\alpha_{1}}}))}}
\times
\end{equation}
$$
\left\{1+\eta'_{1}
\sum_{
\stackrel{i_{\alpha_{1}}'< j_{\alpha_{1}}'< k_{\alpha_{1}}'}
{\left\{i_{\alpha_{1}}',j_{\alpha_{1}}',
k_{\alpha_{1}}'\right\}\in J_{X'_{i'}}}}
t'_{i_{\alpha_{1}}'}t'_{j_{\alpha_{1}}'}
t'_{k_{\alpha_{1}}'}
{\bf 1}_{(w(t'_{i'_{\alpha_{1}}})=w(t'_{j'_{\alpha_{1}}})
=w(t'_{k'_{\alpha_{1}}}))}+
\right.
$$
$$
+\left.
\eta'_{2}
\sum_
{\stackrel{i_{\alpha_{1}}'< j_{\alpha_{1}}'}
{\left\{i_{\alpha_{1}}',j_{\alpha_{1}}'\right\}\in J_{X'_{i'}}}}
t'_{i_{\alpha_{1}}'}t'_{j_{\alpha_{1}}'}
{\bf 1}_{(w(t'_{i'_{\alpha_{1}}})=w(t'_{j'_{\alpha_{1}}}))}
+\eta'_{3}\sum_{i_{\alpha_{1}}'\in J_{X'_{i'}}}
t'_{i_{\alpha_{1}}'}\right\}+r'.
$$
{\bf Here}
\begin{eqnarray}
     \xi' & = & L^{\beta}\xi+\xi'_2  \\
 \xi'_{2} & = & \gamma _{1}\lambda-\gamma _{2}\lambda^2+O(\lambda^3)\\
 \lambda' & = & L^{(2\beta-d)}\lambda-\beta_{1}\lambda^{2}+O(\lambda^3) \\
\eta'_{1} & = & \eta_{1}A+\eta \lambda^2 \\
  \eta'_2 & = & \eta_1B+L^{(2\beta-d)}\eta_2 \\
  \eta'_3 & = & \eta_1C+\eta_2\gamma_{1}+L^{\beta}\eta_{3} \\
      r'& \sim & O\left(\lambda^3\right).
\end{eqnarray}

{\bf Proof}. From our initial hypothesis, follows
\begin{equation}
U'(w')=
\left\langle U_{\xi}(w)\right\rangle_{w'}
\left\langle U_{\lambda}(w)\right\rangle_{w'},
\end{equation}
to errors $O(\xi\lambda)$ at each site. Here we introduce $\xi=O(\lambda^2)$,
so this can be put in $r$. Besides the subindices $\xi$ and $\lambda $
identify the two factors in eq(21). We know, from theorem 3, the trajectory
of $\left\langle U_{\xi}\left(w\right)\right\rangle_{w'}$, so we only have to
study the trajectory of
$\left\langle U_{\lambda}\left(w\right)\right\rangle_{w'}$.\

We split the product in the interaction term
$\left\langle U_{\lambda}\left(w\right)\right\rangle_{w'}$
into two parts and identify
$
\prod_{X'_{i'}\in {\it G}/{\it G_1}}
\mbox{ with }
\prod_{X'_{i'}\in {\it G}}
$
, therefore;
\begin{equation}
\left\langle U_{\lambda}\left(w\right)\right\rangle_{w'}
=\prod_{X'_{i'}\in {\it G}}
\left\langle U_{\lambda}\left(w\right)
\right\rangle^{\left({\it G}_{1}\right)_{X'_{i'}}}_{w'}
\;\mbox{ where }
\end{equation}
\begin{equation}
\left\langle U_{\lambda}\left(w\right)
\right\rangle^{\left({\it G}_{1}\right)_{X'_{i'}}}_{w'}=
\left\{
\prod_{j'\in J_{X'_{i'}}}
e^{q_{1}(L^d-1)r(L^{\beta}t'_{j'})}
\right\}^{-1}\times
\end{equation}
$$
\left\{
\sum_{X_{i}\in ({\it G}_{1})_{X'_{i'}}}\;
\sum_{n_{i'}}\int\prod_{i\in I_{X'_{i'}}}dt_{i}
\right.
\prod_{j'\in J_{X'_{i'}}}
\;\delta (\sum^{m_{j'}}_{{i}=m_{j'-1}+1}\;\;t_{i}-L^{\beta }t'_{j'})
\prod^{m_{j'}}_{{i}=m_{j'-1}+1}
\chi (X_{i}\in LX'_{j'})
$$
$$
\times
\prod^{n-1}_{i=0}
q(X_{i+1}-X_i)
\left.
r^{n_{i'}}\;
e^{-\lambda\sum_{X_{i}\in ({\it G_1})_{X'_{i'}}}
\sum _{i_{\alpha_{1}}< j_{\alpha_{1}}\in J_{X_{i}}}\;\;
t_{i_{\alpha_{1}}}t_{j_{\alpha_{1}}}
{\bf 1}_{(w(t_{i_{\alpha_{1}}})=w(t_{j_{\alpha_{1}}}))}}
\right\}
$$
Recall that $w'$ is fixed; i.e. the walk, after the
renormalization-group map is applied, visits each site
$X'_{i'}\in {\it G}$ a fixed number
of times $n'^{*}_{i'}$. Let us assume $w'$ is such that
$\left\{n'^{*}_{i'}\right\}>0$, and at least once
$n'^{*}_{i'}=1,2,3$ on ${\it G}$. See example in Figure 2. We ask for this
condition to hold, in order to learn about the flow of the interaction
constant in the (self-) intersection of 2 and 3 random walks. In other words,
we ask for a fixed and not totally arbitrary topology for the renormalized
random walk $w'$ on ${\it G}$. To make this condition explicit we rewrite
eq(31) as;
\begin{equation}
\left\langle U_{\lambda}\left(w\right)\right\rangle_{w'}
=\prod_{X'_{i'}\in {\it G}}
\prod_{\left(n'^{*}_{i'}\right)^{k}_{i'=0}}
\left\langle U_{\lambda}\left(w\right)
\right\rangle^{\left({\it G}_{1}\right)_{X'_{i'}}}
_{w'_{n'^{*}_{i'}}}
\end{equation}
Here,
$
\left\langle U_{\lambda}\left(w\right)
\right\rangle^{\left({\it G}_{1}\right)_{X'_{i'}}}
_{w'_{n'^{*}_{i'}}}
$
is the renormalized interaction energy for all possible topologies of walks
$w$ inside the ${\left({\it G}_{1}\right)_{X'_{i'}}}$ coset that, once the
renormalization-group map is applied, corresponds to a fixed topology in
$LX'_{i'}\in {\it G}$.\

Let us introduce a formal Taylor series expansion in $\lambda$, then;
\begin{equation}
\left\langle U_{\lambda}\left(w\right)
\right\rangle^{\left({\it G}_{1}\right)_{X'_{i'}}}
_{w'_{n'^{*}_{i'}}}=
\sum^{\infty}_{s=1}\frac{\left(-\lambda \right)^s}{s!}
\end{equation}
$$
\left\langle
\sum_{X_i\in {({\it G}_1)}_{X'_{i'}}}\;
\sum_
{\stackrel{i_{\alpha_{1}}< j_{\alpha_{1}}<;...;
<i_{\alpha_{s}}< j_{\alpha_{s}}}
{\left\{i_{\alpha_{1}},...,i_{\alpha_{s}}
; j_{\alpha_{1}},...,j_{\alpha_{s}}\right\}
\in J_{X_{i}}}}
t_{i_{\alpha_{1}}}t_{j_{\alpha_{1}}}
{\bf 1}_{(w(t_{i_{\alpha_{1}}})=w(t_{j_{\alpha_{1}}}))}
\times ...
\right.
$$
$$
\left.
...\times t_{i_{\alpha_{s}}}t_{j_{\alpha_{s}}}
{\bf 1}_{(w(t_{i_{\alpha_{s}}})=w(t_{j_{\alpha_{s}}}))}
\chi\left(\left(i_{\alpha_{1}},...,i_{\alpha_{s}};
j_{\alpha_{1}},...,j_{\alpha_{s}}\right)\in (i'_{\alpha_{1}},...,
i'_{\alpha_{n'^{*}_{i'}}})
\right)
\right\rangle ^{l.c.}_{w'}
$$
In this formal series expansion, we are writing
$
\left\langle U_{\lambda}\left(w\right)
\right\rangle^{\left({\it G}_{1}\right)_{X'_{i'}}}
_{w'_{n'^{*}_{i'}}}
$
in terms af all possible classes of topology for walks $w$ inside the
${\left({\it G}_{1}\right)_{X'_{i'}}}$ coset. Each class is an element
of this Taylor series and corresponds to a fix
number $s$ of double (self)-intersections, weighted by $\lambda^s$. Here, the
superscript $l.c.$ means linear contribution. We take into account only
linear contributions to conditional expectations. This approach is considered
to avoid double-counting sites in
walks.\

 To start with, we analyze explicitly the case $n'^{*}_{i'}=1$,
(for some $0\leq i'\leq k$) up to 2nd order in $\lambda$;
\begin{equation}
\left\langle U_{\lambda}\left(w\right)
\right\rangle^{\left({\it G}_{1}\right)_{X'_{i'}}}
_{w'_{n'^{*}_{i'}=1}}=
1-\lambda\times
\end{equation}
$$
\left\langle
\sum_{X_i\in {({\it G}_1)}_{X'_{i'}}}\;
\sum_
{\stackrel{i_{\alpha_{1}}< j_{\alpha_{1}}}
{\left\{i_{\alpha_{1}},
j_{\alpha_{1}}\right\}
\in J_{X_{i}}}}
t_{i_{\alpha_{1}}}t_{j_{\alpha_{1}}}
\left. {\bf 1}_{(w(t_{i_{\alpha_{1}}})=w(t_{j_{\alpha_{1}}}))}
\chi\left(\left(i_{\alpha_{1}},
j_{\alpha_{1}}\right)\in i'_{\alpha{1}}
\right)
\right\rangle ^{l.c.}_{w'}
\right.
$$
$$
+\frac{\lambda^{2}}
{2}
\left\langle
\sum_{\left(X_{i}\in {\it G}_{1}\right)_{X'_{i'}}}\;
\sum_
{\stackrel{i_{\alpha_{1}}< j_{\alpha_{1}}<
i_{\alpha_{2}}< j_{\alpha_{2}}}
{\left\{i_{\alpha_{1}},i_{\alpha_{2}},
j_{\alpha_{1}},j_{\alpha_{2}}\right\}
\in J_{\left(X_{i}\right)}}}
t_{i_{\alpha_{1}}}t_{j_{\alpha_{1}}}\times
\right.
$$
$$
\times\left.
{\bf 1}_{(w(t_{i_{\alpha_{1}}})=w(t_{j_{\alpha_{1}}}))}
 t_{i_{\alpha_{2}}}t_{j_{\alpha_{2}}}
{\bf 1}_{(w(t_{i_{\alpha_{2}}})=w(t_{j_{\alpha_{2}}}))}
\chi\left(\left(i_{\alpha_{1}},i_{\alpha_{2}},
j_{\alpha_{1}},j_{\alpha_{2}}\right)\in i'_{\alpha_{1}}
\right)
\right\rangle ^{l.c.}_{w'}+ r'_{1},
$$
where $r'_1\sim O(\lambda^3)$.\

Thus, eq(35) is written as
\begin{equation}
\left\langle U_{\lambda}\left(w\right)
\right\rangle^{\left({\it G}_{1}\right)_{X'_{i'}},l.c.}
_{w'_{n'^{*}_{i'}=1}}=
1-\left(\gamma_{1}\lambda-\gamma_{2}\lambda^{2}
+O(\lambda^3)\right)
\sum_{i_{\alpha_{1}}'\in J_{X'_{i'}}}
t'_{i_{\alpha_{1}}'}
\end{equation}
$$
\cong e^{-\xi'_{2}
\sum_{i_{\alpha_{1}}'\in J_{X'_{i'}}}
t'_{i_{\alpha_{1}}'}}.
$$
where $\xi'_{2}$ is given as in eq(24),
\begin{equation}
\gamma_{1}=
\left\langle
\sum_{X_i\in ({\it G}_1)_{X'_{i'}}}
\sum_
{\stackrel{i_{\alpha_{1}}< j_{\alpha_{1}}}
{\left\{i_{\alpha_{1}},j_{\alpha_{1}}\right\}\in J_{X_{i}}}}
t_{i_{\alpha_{1}}}t_{j_{\alpha_{1}}}
{\bf 1}_{(w(t_{i_{\alpha_{1}}})=w(t_{j_{\alpha_{1}}}))}
\chi\left(\left(i_{\alpha_{1}},j_{\alpha_{1}}\right)\in i'_{\alpha_{1}}
\right)
\right\rangle ^{l.c.}_{w'}\;\mbox{ and}
\end{equation}
\begin{equation}
\gamma_{2}=
\frac{1}{2}
\left\langle
\sum_{X_i\in ({\it G}_1)_{X'_{i'}}}
\sum_
{\stackrel{i_{\alpha_{1}}< j_{\alpha_{1}}<
i_{\alpha_{2}}< j_{\alpha_{2}}}
{\left\{i_{\alpha_{1}},j_{\alpha_{1}},
i_{\alpha_{2}},j_{\alpha_{2}}\right\}
\in J_{X_{i}}}}
t_{i_{\alpha_{1}}}t_{j_{\alpha_{1}}}
{\bf 1}_{(w(t_{i_{\alpha_{1}}})=w(t_{j_{\alpha_{1}}}))}\times
\right.
\end{equation}
$$
\left.\times
t_{i_{\alpha_{2}}}t_{j_{\alpha_{2}}}
{\bf 1}_{(w(t_{i_{\alpha_{2}}})=w(t_{j_{\alpha_{2}}}))}
\chi\left(\left(i_{\alpha_{1}},j_{\alpha_{1}},
i_{\alpha_{2}},j_{\alpha_{2}}\right)\in i'_{\alpha_{1}}
\right)
\right\rangle ^{l.c.}_{w'}.
$$
$\gamma_1$ and $\gamma_2$ are explained in Figure 3.\

In the same spirit we study the case $n'^{\ast}_{i'}=2$, (for some
$0\leq i'\leq k$) up to second order in $\lambda $;
\begin{equation}
\left\langle U_{\lambda}\left(w\right)
\right\rangle^{\left({\it G}_{1}\right)_{X'_{i'}}lc}
_{w'_{n'^{*}_{i'}=2}}=1-\left(
L^{(2\beta-d)}\lambda-\beta_{1}\lambda^{2}+O(\lambda^3)\right)\times
\end{equation}
$$
\times
\sum_
{\stackrel{i_{\alpha_{1}}'< j_{\alpha_{1}}'}
{\left\{i_{\alpha_{1}}',j_{\alpha_{1}}'\right\}\in J_{X'_{i'}}}}
t'_{i_{\alpha_{1}}'}t'_{j_{\alpha_{1}}'}
{\bf 1}_{(w(t'_{i'_{\alpha_{1}}})=w(t'_{j'_{\alpha_{1}}}))}
$$
$$
\cong
e^{-
\lambda'
\sum_
{\stackrel{i_{\alpha_{1}}'< j_{\alpha_{1}}'}
{\left\{i_{\alpha_{1}}',j_{\alpha_{1}}'\right\}\in J_{X'_{i'}}}}
t'_{i_{\alpha_{1}}'}t'_{j_{\alpha_{1}}'}
{\bf 1}_{(w(t'_{i'_{\alpha_{1}}})=w(t'_{j'_{\alpha_{1}}}))}}.
$$
where $\lambda'$ is given by eq(25) and
\begin{equation}
\beta_1=\frac{1}{2}
\left\langle
\sum_{X_i\in ({\it G}_1)_{X'_{i'}}}
\sum_
{\stackrel{i_{\alpha_{1}}< j_{\alpha_{1}}<
i_{\alpha_{2}}< j_{\alpha_{2}}}
{\left\{i_{\alpha_{1}},j_{\alpha_{1}},
i_{\alpha_{2}},j_{\alpha_{2}}\right\}
\in J_{X_{i}}}}
t_{i_{\alpha_{1}}}t_{j_{\alpha_{1}}}
{\bf 1}_{(w(t_{i_{\alpha_{1}}})=w(t_{j_{\alpha_{1}}}))}
\right. \times
\end{equation}
$$
\times\left.
t_{i_{\alpha_{2}}}t_{j_{\alpha_{2}}}
{\bf 1}_{(w(t_{i_{\alpha_{2}}})=w(t_{j_{\alpha_{2}}}))}
\chi\left(\left(i_{\alpha_{1}},j_{\alpha_{1}},
i_{\alpha_{2}},j_{\alpha_{2}}\right)\in \left(i'_{\alpha_{1}}<
j'_{\alpha_{1}}\right)
\right)
\right\rangle ^{l.c.}_{w'}.
$$
$\beta_1$ is explained in Figure 3.\

Finally, let us present the factor $n'^{*}_{i'}=3$,
(for some $0\leq i'\leq k$), up to 2nd order in $\lambda$;
\begin{equation}
\left\langle U_{\lambda}\left(w\right)
\right\rangle^{\left({\it G}_{1}\right)_{X'_{i'}},l.c.}
_{w'_{n'^{*}_{i'}=3}}=
\left(1+
\right.
\end{equation}
$$
+\eta\lambda^{2}
\sum_{
\stackrel{i_{\alpha_{1}}'< j_{\alpha_{1}}'< k_{\alpha_{1}}'}
{\left\{i_{\alpha_{1}}',j_{\alpha_{1}}',
k_{\alpha_{1}}'\right\}\in J_{X'_{i'}}}}
t'_{i_{\alpha_{1}}'}t'_{j_{\alpha_{1}}'}
t'_{k_{\alpha_{1}}'}
{\bf 1}_{(w(t'_{i'_{\alpha_{1}}})=w(t'_{j'_{\alpha_{1}}})
=w(t'_{k'_{\alpha_{1}}}))}
\left.
+r'_{3}\right)
$$
where $\eta$ is

\begin{equation}
\eta=\frac{1}{2}
\left\langle
\sum_{X_{i}\in ({\it G}_{1})_{X'_{i'}}}
\sum_
{\stackrel{i_{\alpha_{1}}< j_{\alpha_{1}}<
i_{\alpha_{2}}< j_{\alpha_{2}}}
{\left\{i_{\alpha_{1}},j_{\alpha_{1}},
i_{\alpha_{2}},j_{\alpha_{2}}\right\}
\in J_{X_{i}}}}
t_{i_{\alpha_{1}}}t_{j_{\alpha_{1}}}
{\bf 1}_{(w(t_{i_{\alpha_{1}}})=w(t_{j_{\alpha_{1}}}))}
\right. \times
\end{equation}
$$
\times\left.
t_{i_{\alpha_{2}}}t_{j_{\alpha_{2}}}
{\bf 1}_{(w(t_{i_{\alpha_{2}}})=w(t_{j_{\alpha_{2}}}))}
\chi\left(\left(i_{\alpha_{1}},j_{\alpha_{1}},
i_{\alpha_{2}},j_{\alpha_{2}}\right)\in \left(i'_{\alpha_{1}}<
j'_{\alpha_{1}}<k'_{\alpha_{1}}\right)
\right)
\right\rangle ^{l.c.}_{w'} .
$$
$\eta$ is explained in Figure 3.\

Note that in eq(41)

{\footnotesize \begin{equation}
\left\langle
\sum_{X_{i}\in ({\it G}_{1})_{X'_{i'}}}
\sum_
{\stackrel{i_{\alpha_{1}}< j_{\alpha_{1}}}
{\left\{i_{\alpha_{1}},
j_{\alpha_{1}}\right\}
\in J_{X_{i}}}}
t_{i_{\alpha_{1}}}t_{j_{\alpha_{1}}}
{\bf 1}_{(w(t_{i_{\alpha_{1}}})=w(t_{j_{\alpha_{1}}}))}
\chi\left(\left(i_{\alpha_{1}},
j_{\alpha_{1}}\right)\in \left(i'_{\alpha_{1}}< j'_{\alpha_{1}}
< k'_{\alpha_{1}}\right)\right)
\right\rangle ^{l.c.}_{w'}=0
\end{equation}}
for all set of walks $\{w\}\in \Gamma_{n}$, being $\Gamma_n$ the space of
random walks with  only double (self-)intersecting walks.\

Note that the case $n'^{*}_{i'}=1$, for some $0\leq i'\leq k$,
shows how the weakly (self-)avoiding random walk on the
$\left({\it G}_{1}\right)_{X'_{i'}}$ coset can lead to a ``mass"
term; i.e. a local time contribution, after the transformation is
applied. This shall be used in next section to obtain the asymptotic end-to-end
distance of a weakly SARW on a hierarchical lattice, $d=4$, heuristically.
Similarly, the case $n'^{*}_{j'}=3$, for some
$j'\neq i'$ and $0\leq j'\leq k$, shows how a different
realization on the ${\it G}_{1}$ cosets of the weakly SARW can render, a pair
of double intersections of random walks inside the $\left({\it G}_{1}\right)
_{X'_{j'}}$ coset, to the triple intersection of renormalized walks in
$X'_{j'}$ after the renormalization-group map is applied. See Figure 2.\

Writing together cases $n'^{*}_{i'}=1,2,3$ we obtain
\begin{equation}
U'(w')=
\prod_{X'_{i'}\in{\it G}}
\end{equation}
$$
e^{
-\left(L^{\beta}\xi+\xi'_{2}\right)
\sum_{i_{\alpha_{1}}'\in J_{X'_{i'}}}
t'_{i_{\alpha_{1}}'}-
\lambda'
\sum_
{\stackrel{i_{\alpha_{1}}'< j_{\alpha_{1}}'}
{\left\{i_{\alpha_{1}}',j_{\alpha_{1}}'\right\}\in J_{X'_{i'}}}}
t'_{i_{\alpha_{1}}'}t'_{j_{\alpha_{1}}'}
{\bf 1}_{(w(t'_{i'_{\alpha_{1} } })
=w(t'_{j'_{\alpha_{1}} })) } }\times
$$
$$
\left\{1+\eta\lambda^{2}
\sum_{
\stackrel{i_{\alpha_{1}}'< j_{\alpha_{1}}'< k_{\alpha_{1}}'}
{\left\{i_{\alpha_{1}}',j_{\alpha_{1}}',
k_{\alpha_{1}}'\right\}\in J_{X'_{i'}}}}
t'_{i_{\alpha_{1}}'}t'_{j_{\alpha_{1}}'}
t'_{k_{\alpha_{1}}'}
{\bf 1}_{(w(t'_{i'_{\alpha_{1}}})=w(t'_{j'_{\alpha_{1}}})
=w(t'_{k'_{\alpha_{1}}}))}
\right\}+r'.
$$
where $r'\sim O(\lambda^{3})$ bounds the cosets where
$n'^{\ast}_{i'}\geq 4$, for some
$0\leq i'\leq k$. Note that $\lambda^3$ is the leading
contribution to $r'$ if
$n'^{\ast }_{i'}= 4$. As $n'^{\ast }_{i'}$
increases the leading contribution decreases. Here
$$
\xi'=\xi'_{2}+L^{\beta}\xi
$$
$$
\eta'_1=\eta\lambda^2\;\;\;\;\;\;\;
\lambda'=L^{(2\beta-d)}\lambda-\beta'\lambda^2+O(\lambda^3)
$$
We apply once more the renormalization-group map to eq(44). In sake of
clarity let us supress primes in eq(44), so the primed terms always
correspond to the renormalized ones. From the initial hypothesis, approaching
the three walks intersection factor in eq(44) to an exponential, factorize
this as done in eq(33), then expanding the exponential up to the first order,
i.e. $\sim \left(\eta\lambda^2\right)$;
from the result in theorem 3, eq(36), eq(39) and eq(41), follows
\begin{equation}
U'(w')=
\end{equation}
$$
\prod_{X'_{i'}\in{\it G}'}
e^{
-\xi'
\sum_{i_{\alpha_{1}}'\in J_{X'_{i'}}}
t'_{i_{\alpha_{1}}'}-
\lambda'
\sum_
{\stackrel{i_{\alpha_{1}}'< j_{\alpha_{1}}'}
{\left\{i_{\alpha_{1}}',j_{\alpha_{1}}'\right\}\in J_{X'_{i'}}}}
t'_{i_{\alpha_{1}}'}t'_{j_{\alpha_{1}}'}
{\bf 1}_{(w(t'_{i'_{\alpha_{1}}})
=w(t'_{j'_{\alpha_{1}}}))}}\times
$$
$$
\left\{1+\eta'_{1}
\sum_{
\stackrel{i_{\alpha_{1}}'< j_{\alpha_{1}}'< k_{\alpha_{1}}'}
{\left\{i_{\alpha_{1}}',j_{\alpha_{1}}',
k_{\alpha_{1}}'\right\}\in J_{X'_{i'}}}}
t'_{i_{\alpha_{1}}'}t'_{j_{\alpha_{1}}'}
t'_{k_{\alpha_{1}}'}
{\bf 1}_{(w(t'_{i'_{\alpha_{1}}})=w(t'_{j'_{\alpha_{1}}})
=w(t'_{k'_{\alpha_{1}}}))}
\right.
$$
$$
+\left.
\eta'_{2}
\sum_
{\stackrel{i_{\alpha_{1}}'< j_{\alpha_{1}}'}
{\left\{i_{\alpha_{1}}',j_{\alpha_{1}}'\right\}\in J_{X'_{i'}}}}
t'_{i_{\alpha_{1}}'}t'_{j_{\alpha_{1}}'}
{\bf 1}_{(w(t'_{i'_{\alpha_{1}}})
=w(t'_{j'_{\alpha_{1}}}))}
+\eta'_{3}\sum_{i_{\alpha_{1}}'\in J_{X'_{i'}}}
t'_{i_{\alpha_{1}}'}\right\}+r'.
$$
where $r'\sim O(\lambda^3)$ and
$$
\xi'_{2}= \gamma_{1}\lambda-\gamma_{2}\lambda^2
+O(\lambda^3),
$$
$$
\xi'=L^{\beta}\xi+\xi'_2,
$$
$$
\lambda'=L^{(2\beta -d)}\lambda-\beta_{1}\lambda^2+
O(\lambda^3);
\;\;\;
\eta'_{1}=\eta_{1}A+\eta\lambda^{2}
$$
$$
\eta'_{2}=\eta_{1}B,
\;\;\;\; \eta'_{3}=\eta_{1}C
$$
In eq(45) $A$, $B$ and $C$ are  respectively given as follows;
\begin{equation}
A=\left\langle
\sum_{X_i\in ({\it G}_1)_{X'_{i'}}}
\sum_
{\stackrel{i_{\alpha_{1}}< j_{\alpha_{1}}< k_{\alpha_{1}}}
{\left\{i_{\alpha_{1}},j_{\alpha_{1}},k_{\alpha_{1}}\right\}
\in J_{X_{i}}}}
t_{i_{\alpha_{1}}}t_{j_{\alpha_{1}}}
t_{k_{\alpha_{1}}}
{\bf 1}_{(w(t_{i_{\alpha_{1}}})=w(t_{j_{\alpha_{1}}})
=w(t_{k_{\alpha_{1}}}))}
\right. \times
\end{equation}
$$
\times\left.
\chi\left(\left(i_{\alpha_{1}},j_{\alpha_{1}},k_{\alpha_{1}}\right)
\in \left(i'_{\alpha_{1}},j'_{\alpha_{1}},k'_{\alpha_{1}}\right)
\right)
\right\rangle ^{l.c.}_{w'},
$$
\begin{equation}
B=\left\langle
\sum_{X_{i}\in ({\it G}_{1})_{X'_{i'}}}
\sum_
{\stackrel{i_{\alpha_{1}}< j_{\alpha_{1}}< k_{\alpha_{1}}}
{\left\{i_{\alpha_{1}},j_{\alpha_{1}},k_{\alpha_{1}}\right\}
\in J_{X_{i}}}}
t_{i_{\alpha_{1}}}t_{j_{\alpha_{1}}}
t_{k_{\alpha_{1}}}
{\bf 1}_{(w(t_{i_{\alpha_{1}}})=w(t_{j_{\alpha_{1}}})
=w(t_{k_{\alpha_{1}}}))}
\right. \times
\end{equation}
$$
\times\left.
\chi\left(\left(i_{\alpha_{1}},j_{\alpha_{1}},k_{\alpha_{1}}\right)
\in \left(i'_{\alpha_{1}}<j'_{\alpha_{1}}\right)
\right)
\right\rangle ^{l.c.}_{w'} \;\;{\mbox and}
$$
\begin{equation}
C=\left\langle
\sum_{X_{i}\in ({\it G}_{1})_{X'_{i'}}}
\sum_
{\stackrel{i_{\alpha_{1}}<j_{\alpha_{1}}< k_{\alpha_{1}}}
{\left\{i_{\alpha_{1}},j_{\alpha_{1}},k_{\alpha_{1}}\right\}
\in J_{X_{i}}}}
t_{i_{\alpha_{1}}}t_{j_{\alpha_{1}}}
t_{k_{\alpha_{1}}}
{\bf 1}_{(w(t_{i_{\alpha_{1}}})=w(t_{j_{\alpha_{1}}})
=w(t_{k_{\alpha_{1}}}))}
\right. \times
\end{equation}
$$
\times\left.
\chi\left(\left(i_{\alpha_{1}},j_{\alpha_{1}},k_{\alpha_{1}}\right)
\in \left(i'_{\alpha_{1}}\right)
\right)
\right\rangle ^{l.c.}_{w'}.
$$
$A$, $B$ and $C$ are explained in Figure 3.\

Note that, to get the result in eq(45), we have studied the
trajectory, under the renormalization-group map R, of
\begin{equation}
\left\{
\prod_{X_{i}\in ({\it G}_{1})_{X'_{i'}}}\;
e^{
\eta'_1
\sum_{
\stackrel{i_{\alpha_{1}}< j_{\alpha_{1}}< k_{\alpha_{1}}}
{\left\{i_{\alpha_{1}},j_{\alpha_{1}},
k_{\alpha_{1}}\right\}\in J_{X_{i}}}}
t_{i_{\alpha_{1}}}t_{j_{\alpha_{1}}}
t_{k_{\alpha_{1}}}
{\bf 1}_{(w(t_{i_{\alpha_{1}}})=w(t_{j_{\alpha_{1}}})
=w(t_{k_{\alpha_{1}}}))}
}
\right\}.
\end{equation}
When we apply the renormalization-group map to eq(49) we end up with the
contributions in $A,B$ and $C$. In other words, we use the same procedure
above developed for the double (self-)intersection of random walks but for
the triple (self-)intersection of random walks up to order $\lambda^{2}$,
this corresponds to the first term in the corresponding Taylor
series expansion.\

Applying the renormalization-group map to eq(45) we obtain eq(22) (eq(44) and
eq(45) are particular cases of eq(22)). Then, the proof follows from
induction. We apply the renormalization-group map to $U'(w')$ (eq(22)).
Recall that we supress primes in eq(22), thus primed terms are the
renormalized ones. From our original hypotheses and due to the hierarchical
structure of the lattice;
\begin{equation}
U'(w')=
\end{equation}
$$
\prod_{X'_{i'}\in{\it G}}
\left\langle
\prod_{X_{i}\in ({\it G_1})_{X'_{i'}}}
e^{
-\xi
\sum_{i_{\alpha_{1}}\in J_{X_{i}}}
t_{i_{\alpha_{1}}}}
\right\rangle^{l.c}_{w'}\times
$$\

$$
\times
\left\langle
\prod_{X_{i}\in ({\it G_1})_{X'_{i'}}}
e^{
-\lambda
\sum_
{\stackrel{i_{\alpha_{1}}< j_{\alpha_{1}}}
{\left\{i_{\alpha_{1}},j_{\alpha_{1}}\right\}\in J_{X_{i}}}}
t_{i_{\alpha_{1}}}t_{j_{\alpha_{1}}}
{\bf 1}_{(w(t_{i_{\alpha_{1}}})=w(t_{j_{\alpha_{1}}}))}}
\right\rangle^{l.c}_{w'}\times
$$\

$$
\left\langle1+\eta_{1}
\sum_{X_{i}\in ({\it G_1})_{X'_{i'}}}
\sum_{
\stackrel{i_{\alpha_{1}}< j_{\alpha_{1}}< k_{\alpha_{1}}}
{\left\{i_{\alpha_{1}},j_{\alpha_{1}},
k_{\alpha_{1}}\right\}\in J_{X_{i}}}}
t_{i_{\alpha_{1}}}t_{j_{\alpha_{1}}}
t_{k_{\alpha_{1}}}
{\bf 1}_{(w(t_{i_{\alpha_{1}}})=w(t_{j_{\alpha_{1}}})
=w(t_{k_{\alpha_{1}}}))}+
\right.
$$\

$$
+\eta_{2}
\sum_{X_{i}\in ({\it G_1})_{X'_{i'}}}
\sum_
{\stackrel{i_{\alpha_{1}}< j_{\alpha_{1}}}
{\left\{i_{\alpha_{1}},j_{\alpha_{1}}\right\}\in J_{X_{i}}}}
t_{i_{\alpha_{1}}}t_{j_{\alpha_{1}}}
{\bf 1}_{(w(t_{i_{\alpha_{1}}})=w(t_{j_{\alpha_{1}}}))}+
$$
$$
\left.
+\eta_{3}
\sum_{X_{i}\in ({\it G_1})_{X'_{i'}}}
\sum_{i_{\alpha_{1}}\in J_{X_{i}}}
t_{i_{\alpha_{1}}}\right\rangle^{l.c}_{w'}
+r'.
$$
Assuming ${n^{\ast}}'_{i'}=1,2,3$ and carrying out a Taylor
series expansion up to order $\lambda^{2}$, it is straightforward to
prove that eq(50) leads to eq(22), provided we use the same bookkeeping
device above explained. We just need
to apply theorem 3, eq(36), eq(39), eq(41) and
\begin{equation}
\left\langle
\prod_{X_{i}\in ({\it G_1})_{X'_{i'}}}
exp(
\eta_{1}
\sum_{
\stackrel{i_{\alpha_{1}}< j_{\alpha_{1}}< k_{\alpha_{1}}}
{\left\{i_{\alpha_{1}},j_{\alpha_{1}},
k_{\alpha_{1}}\right\}\in J_{X_{i}}}}
t_{i_{\alpha_{1}}}t_{j_{\alpha_{1}}}
t_{k_{\alpha_{1}}}
{\bf 1}_{(w(t_{i_{\alpha_{1}}})=w(t_{j_{\alpha_{1}}})
=w(t_{k_{\alpha_{1}}}))}+
\right.
\end{equation}
$$
\left.+
\eta_{2}
\sum_
{\stackrel{i_{\alpha_{1}}< j_{\alpha_{1}}}
{\left\{i_{\alpha_{1}},j_{\alpha_{1}}\right\}\in J_{X_{i}}}}
t_{i_{\alpha_{1}}}t_{j_{\alpha_{1}}}
{\bf 1}_{(w(t_{i_{\alpha_{1}}})=w(t_{j_{\alpha_{1}}}))}
+\eta_{3}
\sum_{i_{\alpha_{1}}\in J_{X_{i}}}
t_{i_{\alpha_{1}}})\right\rangle^{l.c}_{w'}=
$$
$$
\prod_{\left(n'^{*}_{i'}\right)^{n'}_{i'=0}}
\left\langle
\prod_{X_{i}\in ({\it G_1})_{X'_{i'}}}
e^{
\eta_{3}
\sum_{i_{\alpha_{1}}\in J_{X_{i}}}
t_{i_{\alpha_{1}}}}
\right\rangle^{l.c.}_{w'_{n'^{*}_{i'}}}
$$
$$
\times
\left\langle
\prod_{X_{i}\in ({\it G_1})_{X'_{i'}}}
e^{\eta_{1}
\sum_{
\stackrel{i_{\alpha_{1}}< j_{\alpha_{1}}< k_{\alpha_{1}}}
{\left\{i_{\alpha_{1}},j_{\alpha_{1}},
k_{\alpha_{1}}\right\}\in J_{X_{i}}}}
t_{i_{\alpha_{1}}}t_{j_{\alpha_{1}}}
t_{k_{\alpha_{1}}}
{\bf 1}_{(w(t_{i_{\alpha_{1}}})=w(t_{j_{\alpha_{1}}})
=w(t_{k_{\alpha_{1}}}))}}
\right\rangle^{l.c}_{w'_{{n^{*}}'_{i'}}}
$$
$$
\times
\left\langle
\prod_{X_{i}\in ({\it G_1})_{X'_{i'}}}
e^{
\eta_{2}
\sum_
{\stackrel{i_{\alpha_{1}}< j_{\alpha_{1}}}
{\left\{i_{\alpha_{1}},j_{\alpha_{1}}\right\}\in J_{X_{i}}}}
t_{i_{\alpha_{1}}}t_{j_{\alpha_{1}}}
{\bf 1}_{(w(t_{i_{\alpha_{1}}})=w(t_{j_{\alpha_{1}}}))}}
\right\rangle^{l.c.}_{w'_{{n^{*}}'_{i'}}}
=
$$
$$
\left\{1+A\eta_{1}
\sum_{
\stackrel{i_{\alpha_{1}}'< j_{\alpha_{1}}'< k_{\alpha_{1}}'}
{\left\{i_{\alpha_{1}}',j_{\alpha_{1}}',
k_{\alpha_{1}}'\right\}\in J_{X'_{i'}}}}
t'_{i_{\alpha_{1}}'}t'_{j_{\alpha_{1}}'}
t'_{k_{\alpha_{1}}'}
{\bf 1}_{(w(t'_{i'_{\alpha_{1}}})=w(t'_{j'_{\alpha_{1}}})
=w(t'_{k'_{\alpha_{1}}}))}
+
\right.
$$
$$
+\eta'_2
\sum_
{\stackrel{i_{\alpha_{1}}'< j_{\alpha_{1}}'}
{\left\{i_{\alpha_{1}}',j_{\alpha_{1}}'\right\}\in J_{X'_{i'}}}}
t'_{i_{\alpha_{1}}'}t'_{j_{\alpha_{1}}'}
{\bf 1}_{(w(t'_{i'_{\alpha_{1}}})=w(t'_{j'_{\alpha_{1}}}))}
\left.
+\eta'_{3}\sum_{i_{\alpha_{1}}'\in J_{X'_{i'}}}
t'_{i_{\alpha_{1}}'}\right\}+r'_{3}
$$
where $A$, $\eta'_{2}$ and $\eta'_{3}$ are given by Figure 3, eq(27), and
eq(28); $r'\sim O(\lambda^{3})$. Note that $A\eta_1$ in eq(51) plus the
contribution from eq(41) defines $\eta'_{1}$ as done in eq(26). Eq(51) is
written up to $O(\lambda^2)$ terms.\

Q.E.D.\

We claim theorem 4 is the space-time renormalization-group trajectory of the
weakly SARW energy interaction studied by Brydges, Evans and Imbrie \cite{Ev},
provided $\beta=2$ and $d=4$. In reference [5] the trajectory of a
$\lambda\phi^{4}$ superalgebra valued interaction was studied (this can be
understood in terms of intersection of random walks due to the Mc Kane,
Parisi, Sourlas Theorem \cite{Mk}) using a field-theoretical version of the
renormalization-group map. The field theory is defined on the same
hierarchical lattice we are studying here. In this paper, we provide exact
probabilistic expressions for $\lambda'$ and $\xi'$ (which are not given in
reference [5]), these are crucial to propose an heuristic proof for the
asymptotic behavior of the end-to-end distance of a weakly SARW. To do so we
just need to calculate $\gamma_{1}$ from eq (37) and $\beta_{1}$ from eq(40).\

Finally, we summarize important features of our method; \newline
a) the conditional expectation of $U(w)$ can be approached in terms of the
product of conditional expectations.\

b) We take into account only linear contributions to conditional expectations
for probabilities on $\Lambda_{n}$.\

c)  Formal Taylor series expansions are introduced.\

d) We assume our model to be such that each step the renormalization-group
map is applied, the number of times the renormalized walk visits any site of
the lattice is 1, 2 and 3 at least once (i.e. a fixed and not totally
arbitrary topology for the renormalized random walk). See Figure 2.

e) Finally, we take advantage of the hierarchical structure of the lattice.
Since ${\it G}'={\it G}/{\it G}_{1}\approx {\it G}$ and the
map is local, the renormalization-group transformation descends to the
study of walks in the ${\it G}_1$ cosets.\

{}From all of these, we obtain, after applying the
renormalization-group map, the fixed form for a weakly SARW that penalizes,
roughly speaking, the (self-)intersection of two random walks by a factor
($e^{-\lambda}$, $\lambda >0$ and small). Furthermore, this fixed form is the
random walk version (for $d=4$, $\beta=2$) of the one obtained from a
field-theoretical renormalization-group map for a $\lambda \phi^{4}$ model
recently reported by Brydges, Evans and Imbrie \cite{Ev}. We obtain an exact
probabilistic expression for the parameters that appear in the flow of the
interaction factor $\lambda$ which is not given in reference [5]. This shall
be used in next section for the heuristic study of the asymptotic behavior of
the end-to-end distance for a weakly SARW model that punishes the
(self-)intersection of two random walks.

\section{Asymptotic end-to-end distance of a weakly SARW on a hierarchical
lattice in dimension four. An heuristic example as a testing ground.}

The process of renormalizing the lattice is completed by reducing all
dimensions of the new lattice by a factor $L$ each step the
renormalization-group map is applied so we end up with exactly the same
lattice we start with. For a diffusive simple random walk model we reduce
waiting times
by $L^{2}$ each step we apply the renormalization-group transformation.
Moreover, when we iterate probabilities, the end-to-end distance shrinks by
a factor $L$ at each interaction, because in renormalizing the lattice we
divide every length, including the end-to-end distance, by $L$ \cite{Ma}
\cite{Bi}.
{}From this viewpoint we intend to understand, heuristically, the asymptotic
end-to-end distance of a weakly SARW on a hierarchical lattice in $d=4$,
thereby providing a new probabilistic meaning to this magnitude.\

For weakly SARW, we generalize the standard scaling factor for local times of
the renormalization transformation above as described by including, up to
$O(\lambda)$, the contribution of the self-repulsion term to renormalized
local times. Namely, from the renormalization-group map on weakly SARW,
renormalized local times are generated from the interaction. In the field
theoretical approach this corresponds to generating mass. Equivalently, we
can say that the interaction kills the process at a specific rate. If we take
into account only $O(\lambda)$ contributions to this and follow standard
thinking, the well known asymptotic end-to-end distance for the weakly SARW
in $d=4$ follows. By including higher order contributions in $\lambda$ to
renormalized local times (as we have already shown this is not the case for
weakly SARW on the hierarchical lattice, because these contributions are no
significant), and/or different dimension for the lattice, the functional form
of the end-to-end distance changes drastically. Moreover, from our method,
the exponent of the logarithmic correction involved is expressed in terms of
conditional expectations for random walks on the lattice, that upon
calculation, give the well known exponent. In Figure 4 the contributions to
the scaling factor proposed for the weakly SARW used to explain the
asymptotic end-to-end distance in the hierarchical lattice are depicted.\

We remark that the proposition presented in this section involves heuristic
considerations in order to understand, from a probabilistic real-space
viewpoint, the asymptotic end-to-end distance for the weakly SARW on a
hierarchical lattice in $d=4$. This has already been conjectured heuristically
before by other means. Recently a rigorous proof has been given, provided
properties of the Green function are known, in the field theoretical approach
\cite{Im}. Our proposition is anyway presented  as a testing ground for
our method, and for giving probabilistic meaning to the exponent involved
in the logarithmic correction. Once the method shows to be useful for
explaining well known results (at least heurstically), we shall apply
this on more complicated cases, for example kinetically growing measure
model. These are renormalizable, in the field theoretical limit, only for
particular cases. In the process of taking the continuum limit of these
discrete models some memory is lost. Our method is suitable of being applied
on the discrete models. This can be done both, heuristically and rigorously.\

A final remark before introducing the main point of this section is about
the finiteness of moments for random walks on a hierarchical lattice. The
end-to-end distance for the weakly SARW, d=4, on the hierarchical lattice, is
independent of the moment used to obtain it, as should be, provided this is
finite. Let $\left\langle w^{\alpha}(T)\right\rangle$ be an $\alpha$-moment
of the random walk, it is known that the only finite moments for diffusive
random walks on a hierarchical lattice are $0<\alpha<2$ \cite{Im}. This range
of $\alpha$ values is used to obtain the end-to-end distance in the
following \newline

{\bf Proposition. For d=4, up to $O(\lambda)$, the generated renormalized local
times (mass for the field or killing rate for the process), from applying the
renormalization-group map on the interaction, is such that the asymptotic
behavior of the end-to-end distance for a weakly SARW  that penalizes
the intersection of two random walks is $T^{1/2}log^{1/8}T$ as
T tends to infinity}.\

{\bf Proof}. After applying $(p)$ times the renormalization-group
transformation on $\left\langle w^{\alpha}(T)\right\rangle^{1/\alpha}$ we
have
\begin{equation}
\left\langle w^{\alpha}(T)\right\rangle^{1/\alpha}=
\frac{\left\langle w^{\alpha}(1)\right\rangle^{1/\alpha (0)}}{L^{p}}
\end{equation}
where we have chosen a system of units such that, for $p=0$, $T=1$. Hereafter,
$\left\langle w^{\alpha}(1)\right\rangle^{1/\alpha (0)}$=D, constant. Here,
we are following the standard procedure for scaling length type magnitudes
\cite{Bi} ( i.e. $\left\langle w^{\alpha}(T)\right\rangle^{1/\alpha'}$=
$\frac{\left\langle w^{\alpha}(T)\right\rangle^{1/\alpha}}{L}$). Moreover, by
$\left\langle w^{\alpha}(1)\right\rangle^{1/\alpha}$ we mean
$\left\langle w^{\alpha}(1)\right\rangle^{1/\alpha (p)}$. Since in
renormalizing the lattice we divide every length, including the end-to-end
distance by $L$, then, upon p iterations, eq(52) follows. This is exactly
what is done in scaling correlation lengths but used here on the end-to-end
distance, both length type magnitudes. So eq(52) becomes
\begin{equation}
\left\langle w^{\alpha}(T)\right\rangle^{1/\alpha}=L^{-p}D
\end{equation}
On the other hand, from what we stated in theorem 5 we know that
\begin{equation}
T=\frac{1}
{L^{2p}\prod^{p}_{i=1}(1+\gamma^{\ast}_{1}\lambda^{(i)})}\;\;
\mbox{   ,where}
\end{equation}
$\gamma^*_1$=$\gamma_1/L^2$ and by $T$ we mean $T^{(p)}$. Here we have included
up to $O(\lambda)$ contributions to renormalized local times for scaling the
running time of the process. In this scaling factor of the renormalization
transformation, the $O(\lambda)$ contribution comes from the first term in
right-hand side of eq(24). See Figure 4.\

{}From eq(54) and eq(53) follows
\begin{equation}
\left\langle w^{\alpha}(T)\right\rangle^{1/\alpha}=DT^{1/2}
\left(\prod^{p}_{i=1}(1+\gamma^{\ast}_{1}\lambda^{(i)})\right)^{1/2}
\end{equation}
or
\begin{equation}
\left\langle w^{\alpha}(T)\right\rangle^{1/\alpha} \sim
DT^{1/2}\left(e^{\gamma^{\ast}_{1}\sum^{(p)}_{i=1}
\lambda^{(i)}}\right)^{1/2}.
\end{equation}
For $\beta=2$, $d=4$ and up to order $(\lambda^{(i)})^2$, follows
\begin{equation}
\lambda^{(i+1)}=\lambda^{(i)}-\beta_{1}(\lambda^{(i)})^{2}.
\end{equation}
Introducing the solution of the eq(57) recursion into eq(56) this becomes
\begin{equation}
\left\langle w^{\alpha}(T)\right\rangle^{1/\alpha} \sim DT^{1/2}
e^{\frac{\gamma^{\ast}_{1}}{2\beta_{1}}ln p}
\end{equation}
or
\begin{equation}
\left\langle w^{\alpha}(T)\right\rangle^{1/\alpha} \sim DT^{1/2}
(p)^{\frac{\gamma^{\ast}_{1}}{2\beta_{1}}}
\end{equation}
In eq(59) we have assumed p to be large enough so
$\lambda^{-1}<<\beta_{1}(p)$. Taking the asymptotic limit we rewrite eq(59)
as
\begin{equation}
\left\langle w^{\alpha}(T)\right\rangle^{1/\alpha} \sim DT^{1/2}
log^{\frac{\gamma^{\ast}_{1}}{2\beta_{1}}}T,
\end{equation}
which is the asymptotic behavior of  the end-to-end distance.\

It only remains to know the value of
$\left(\frac{\gamma^{\ast}_{1}}{\beta_{1}}\right)$. Actually we can
calculate $\gamma^{\ast}_{1}$ and $\beta_{1}$ from their definitions.\

Let us start with $\gamma_{1}$, from eq(37) we obtain
\begin{equation}
\gamma_{1}=\left\{
\left(
\prod_{i'_{\alpha_{1}}\in J_{X'_{i'}}}
e^{q_{1}(L^d-1)r(L^{\beta}t'_{i'_{\alpha_{1}}})}
\right)^{-1}\times
\right.
\end{equation}
$$
\times
\left(
\sum_{n_{i'}}\;L^{d}\int\prod_{i\in I_{X'_{i'}}}
\;dt_{i}\;\prod_{i'_{\alpha_{1}}\in J_{X'_{i'}}}
\;\delta
(\sum^{m_{i'_{\alpha_{1}}}}_{{i}=m_{i'_{\alpha_{1}}-1}+1}\;\;t_{i}-
L^{\beta }t'_{i'_{\alpha_{1}}})\times \right.
$$
$$
\times
\left(
\begin{array}{c}
n_{i'}+1 \\
2
\end{array}
\right)
(q_{1})^{\left(n_{i'}-1\right)}r^{\left(n_{i'}\right)}
(L^{d}-1)\times...\times
(L^{d}-(n_{i'}-1))
\times
$$
$$
\left.
\times
\sum_{X_{i}\in ({\it G}_{1})_{X'_{i'}}}
\sum_
{\stackrel{i_{\alpha_{1}}< j_{\alpha_{1}}}
{\left\{i_{\alpha_{1}},j_{\alpha_{1}}\right\}\in J_{X_{i}}}}
t_{i_{\alpha_{1}}}t_{j_{\alpha_{1}}}
{\bf 1}_{(w(t_{i_{\alpha_{1}}})=w(t_{j_{\alpha_{1}}}))}
\chi\left(\left(i_{\alpha_{1}},j_{\alpha_{1}}\right)\in i'_{\alpha_{1}}
\right)
\right\}^{l.c.}
$$
Hereafter, we assume $L^{d}>>n_{i'}$, so
$$
\left(L^{d}-(n_{i'}-1)\right)\sim \left(L^{d}-1\right),
$$
i.e. the number of points inside each ${\it G}_{1}$ coset is larger than the
number
of steps the walk $w'$ spends inside each L-block. Thus, the numerator of
eq(61) can be written as
\begin{equation}
\frac{\left(\sum_{i'_{\alpha_{1}}\in J_{X'_{i'}}} L^{\beta}t'
_{i'_{\alpha_{1}}}\right)^{2} }
{2q_{1}(L^{d}-1)}
\prod_{i'_{\alpha_{1}}\in J_{X'_{i'}}}
\int^{1}_{0}\sum_{n_{i'}}\left(n_{i'}+1\right)
\left(n_{i'}\right)^{2}\times
\end{equation}
$$
\times
\left(n_{i'}-1\right)
\frac{ \left( rq_{1}(L^{d}-1)(L^{\beta}t'_{i'_{\alpha_{1} } })
(1-t-t^{\ast})\right)^{n_{i'}}}
{n_{i' }! }
tdtt^{\ast}dt^{\ast}.
$$
where we have taken $(1-t-t^{\ast})^{n_{i'}-2}$ $\sim $
$(1-t-t^{\ast})^{n_{i'}}$. To obtain an asymptotic estimate of eq(61), we
assume that the following holds
$$
\left(n_{i'}+1\right)\left(n_{i'}\right)^{2}
\left(n_{i'}-1\right)\sim
\frac{\left(n_{i'}\right)!}
{\left(n_{i'}-4\right)!}
$$
Althought from this follows $n_{i'}$ chosen to be large, we certainly assume
finite local times after the renormalization-group transformation is
applied.\

Substituting eq(62) in eq(61), with a jumping rate $r$ such that\newline
$rq_{1}(L^{d}-1)\sim 1$ (as done in reference[5]) and for $\beta =2$
we obtain, in the asymptotic limit, $\gamma^{\ast}_{1}\sim 8$ provided
$rq_{1}(L^{d}-1)\sim 1$.\

To calculate $\beta_{1}$ we use eq(40).
\begin{equation}
\beta_{1}=\frac{1}{2}
\left\{
\left(
\prod_{i'_{\alpha_{1}}\in J_{X'_{i'}}}
e^{q_{1}(L^d-1)r(L^{\beta}t'_{i'_{\alpha_{1}}})}
\right)^{-1}\times
\right.
\end{equation}
$$
\times
\left(L^{d}
\sum_{(n_{i'_{a}})}\;
\sum_{(n_{i'_{b}})}\;
\int
\prod_{i_{a}\in I_{X'_{i'}}}
\;dt_{i_{a}}\;
\prod_{i_{b}\in I_{X'_{i'}}}
\;dt_{i_{b}}\;
\right.
\prod_{i'_{\alpha_{1}}\in J_{X'_{i'}}}
\;\delta
(\sum^{m_{i'_{\alpha_{1}}}}_{{i_{a}}=m_{i'_{\alpha_{1}}-1}+1}\;\;
t_{i_{a}}-
L^{\beta }t'_{i'_{\alpha_{1}}})\times
$$
$$
\times
\prod_{j'_{\alpha_{1}}\in J_{X'_{i'}}}
\;\delta
(\sum^{m_{j'_{\alpha_{1}}}}_{{i_{b}}=m_{j'_{\alpha_{1}}-1}+1}\;\;
t_{i_{b}}-
L^{\beta }t'_{j'_{\alpha_{1}}})
\left(
\begin{array}{c}
n_{i'_{a}}+1 \\
2
\end{array}
\right)
\left(
\begin{array}{c}
n_{i'_{b}}+1 \\
2
\end{array}
\right)
(q_{1})^{\left(n_{i'_{a}}\right)}r^{\left(n_{i'_{a}}\right)}\times
$$
$$
\times
(q_{1})^{\left(n_{i'_{b}}-1\right)}r^{\left(n_{i'_{b}}\right)}
(L^{d}-1)\times...\times
(L^{d}-(n_{i'_{a}}+n_{i'_{b}}-1))
\times
$$
$$
\times
\sum_{X_{i}\in ({\it G}_{1})_{X'_{i'}}}
\sum_
{\stackrel{i_{\alpha_{1}}< j_{\alpha_{1}}<
i_{\alpha_{2}}< j_{\alpha_{2}}}
{\left\{i_{\alpha_{1}},j_{\alpha_{1}},
i_{\alpha_{2}},j_{\alpha_{2}}\right\}
\in J_{X_{i}}}}
t_{i_{\alpha_{1}}}t_{j_{\alpha_{1}}}
{\bf 1}_{(w(t_{i_{\alpha_{1}}})=w(t_{j_{\alpha_{1}}})|
 i_{\alpha_{1}},j_{\alpha_{1}}\leq n_{i'_{a}})}
\times
$$
$$
\times
\left.
\left.
t_{i_{\alpha_{2}}}t_{j_{\alpha_{2}}}
{\bf 1}_{(w(t_{i_{\alpha_{2}}})=w(t_{j_{\alpha_{2}}})|
 i_{\alpha_{1}},j_{\alpha_{1}}\leq n_{i'_{b}})}
\chi\left(\left(i_{\alpha_{1}},j_{\alpha_{1}},
i_{\alpha_{2}},j_{\alpha_{2}}\right)\in \left(i'_{\alpha_{1}}<
j'_{\alpha_{1}}\right)
\right)
\right)
\right\}^{l.c.}
$$
As we did for the calculation of $\gamma_{1}$, we assume $L^{d}>>n_{i'}$, so
\begin{equation}
\left(L^{d}-(n_{i'_{a}}-1)\right)\sim (L^{d}-1)\;\mbox{  and  }\;
\left(L^{d}-(n_{i'_{b}})\right)\sim (L^{d}-1).
\end{equation}
Furthermore, we choose
$
n_{i'_{b}}\sim n_{i'_{b}}-1,
$
and approximations in eq(62) to hold for both
$n_{i'_{a}}$ and $n_{i'_{b}}$ with a jumping rate $r$ such that
$rq_{1}(L^{d}-1)\sim 1$. For $\beta=2$ in the asymptotic limit, we
obtain $\beta_{1}\sim 32$ provided $rq_{1}(L^{d}-1)\sim 1$.\

Finally eq(60) becomes
\begin{equation}
\left\langle
w^{\alpha}(T)\right\rangle^{1/\alpha}
\sim (DT)^{1/2}log^{1/8}T
\end{equation}
Q.E.D.\

Note that $d=4$ is the only choice that renders eq(25) (for $\beta=2$)
to a recursion as simple as eq(57) provided that $rq_{1}(L^{d}-1)\sim 1$.\

We want to remark that the heuristic study of the asymptotic end-to-end
distance of a weakly SARW on a hierarchical lattice, $d=4$, is independent of
hypotesis a) in the summary of former section. This is because we could have
obtained the $O(\lambda)$ contribution to renormalized local times without
introducing initial mass into the process.\

\section{Summary}
In this paper we present a real space renormalization-group map,
on the space of probabilities, to study  weakly SARW that penalizes the
(self-)intersection of two random walks for a hierarchical lattice, in
dimension
four. This hierarchical lattice has been labeled by elements of a countable,
abelian group ${\it G}$. For any random function $F(w)$ on ${\it G}$ of the
form described in Section 3, i.e. factorizable on the lattice, ( see eq(17)
and eq(21) for examples of suitable $P(w)$) we can descend
from the study of the space of walks on the whole lattice to the trajectory
in the contracting ${\it G}_{1}$ cosets. Then we show how the L\'{e}vy
process studied in reference [5] is a particular case of the processes that
are (or flow to) fixed points of the renormalization-group map. We apply the
renormalization-group map on some random walk models with configurational
measure, working out explicitly the weakly SARW case. An heuristic proof of
the end-to-end distance for a weakly SARW on a hierarchical lattice is
derived. This gives a new probabilistic meaning to the exponent of the
logarithmic correction.\

In Section 4 we study a weakly SARW that penalizes the (self-)intersection of
two random walks.
The weakly SARW probability studied, involves a factor linear in local times,
i.e. a random walk representation of a field-theoretical
gaussian component that adds to the corresponding term produced for the
renormalization-group map applied on the weakly SARW. We
show how this probability  flows to a fixed form (the random walk version of
the
field-theoretical result given in reference [5]) relying on;\

a) An hypothesis that assumes we can approach the expectation of the
interaction energy in terms of the product of expectations for each of its
factors, conditioned to applying the renormalization-group map.\

b) The hierarchical metric space used to label the lattice that allows,
for $F(w)$ of the form described in Section 3 (f.e. eq(17) and eq(21)), the
factorization in terms of the quotient group ${\it G}/{\it G}_{1}$ and the
image of the renormalization-group map on the cosets ${\it G}_{1}$, each step
the  renormalization-group is applied.\

c) A class of realizations of the model such that, each step we apply
the renormalization-group transformation, the renormalized fixed walk
visits $1, 2, 3$ times different sites in the lattice
${\it G}/{\it G}_{1}$, at least once. Other realizations might not allow us
to study the flow of the (self-)intersecting coefficients that are interesting
for us.\

d) A formal Taylor series expansion in $\lambda$ from which, upon
renormalization, we use only the linear contributions. \

Our result improves the field-theoretical approach \cite{Ev} by obtaining an
exact expression for the parameters that appear in the flow of $\lambda$ and
$\xi$. This is a crucial feature used to obtain heuristically the asymptotic
behavior of the end-to-end distance for a weakly SARW that penalizes the
(self-)intersection of two random walks. Furthermore, the method here
presented is full of physical intuition and suitable of being applied to
discrete kinetically growing measure models.\

Following standard thinking we shrink all space and time magnitudes each step
the map is applied. We shrink time taking into account $O(\lambda)$
contributions to renormalized local times, generated by applying the
renormalization-group map to the weakly SARW that penalizes (self-)
intersections of walks. Length type magnitudes are shrunk as usual. We
present this, as a possible origin for the expression
$\left\langle w^{\alpha}(T)\right\rangle^{1/\alpha}\sim (DT)^{1/2}log^{1/8}T$
as T
tends to infinity, in $d=4$, for a weakly SARW that penalizes the
(self-)intersection of two random walks on a hierarchical lattice.\

\vspace*{10mm}

\section{Acknowledgments}
This work was partially supported by CONACYT REF 4336E9406, M\'{e}xico.
I wish to thank S.N. Evans,  Department of Statistics, UC Berkeley, for
helpful discussion and ITD, UC Davis for its hospitality.

\newpage

\section {Figure Captions}

Figure 1: a) One dimensional and b) two dimensional hierarchical lattices.
Here $,X_k,X_{k-1},...)$ stands for $(...,0,X_k,X_{k-1},...)$.\

\vspace*{2cm}

Figure 2: An example of locality and factorization property of the
renormalization-group map for a fixed, totally arbitrary $w'$. Contributions
to formal Taylor series expansion in the interaction are also depicted.\

\vspace*{2cm}

Figure 3: Conditional expectations of local times involved in renormalized
weakly SARW up to $O(\lambda^2)$.\

\vspace*{2cm}

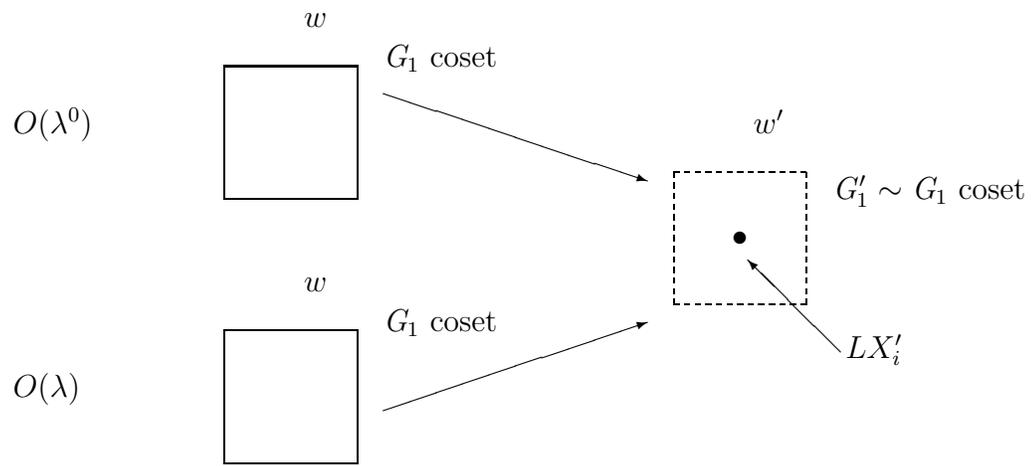
Figure 4: Classes of contributions to the scaling factor for the renormalized
running time of the process up to $O(\lambda)$. From the renormalization-group
map transformation.

\newpage

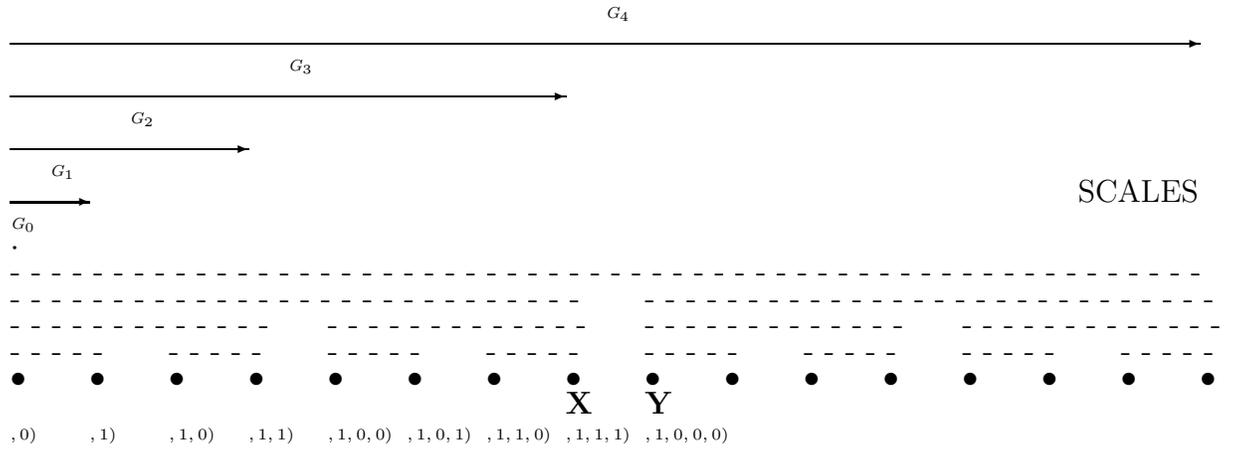
\begin{picture}(350,200)(0,0)
\put(0,155){$O(\lambda^0)$}
\put(80,130){\framebox(50,50)}
\put(110,195){$w$}
\put(140,180){${\it G}_1$ coset}
\put(140,170){\vector(3,-1){100}}
\put(0,55){$O(\lambda)$}
\put(80,30){\framebox(50,50)}
\put(110,95){$w$}
\put(140,80){${\it G}_1$ coset}
\put(140,50){\vector(3,1){100}}
\put(250,90){\dashbox{2}(50,50){$\bullet$}}
\put(280,155){$w'$}
\put(315,70){$LX'_i$}
\put(313,72){\vector(-1,1){35}}
\put(310,130){${\it G}'_1\sim {\it G}_1$ coset}
\put(200,0){Figure 4:}
\end{picture}
\newpage
\begin{tabular}{||l|l|l|l|l|r||}\hline\hline
$\gamma_{1}$                 &
is the                       &
one                          &
			     &
(self-)intersection(s) of walks $w$ &
once\\ \cline{1-1} \cline{3-3}

$\gamma_{2}$                              &
contribution to                           &
two                                       &
double                                    &
 inside the ${\it G}_1$ coset, that after &
\\ \cline{1-1} \cline{6-6}

$\beta_{1}$                    &
the linearized                  &
			       &
			       &
 applying the renormalization  &
twice \\  \cline{1-1} \cline{6-6}

$\eta$                     &
 conditional               &
			   &
			   &
 -group map corresponds to &
three \\ \cline{1-1} \cline{3-3} \cline{4-4}

$A$                                                &
expectation                                        &
one                                                &
						   &
 $LX'_{i'}\in {\it G}\sim {\it G}/{\it G}_1$, given &
times \\ \cline{1-1} \cline{6-6}

$B$                          &
of local times               &
			     &
triple                       &
 a fixed renormalized random &
twice \\ \cline{1-1} \cline{6-6}

$C$                             &
given by            &
				 &
				 &
 walk $w'$ that visits $X'_{i'}$ &
once \\ \hline \hline
\end{tabular}
\vspace*{1cm}
\newline
\hspace*{6cm}Figure 3:

\newpage

\begin{figure}
\begin{picture}(500,500)(0,0)
\put(182,158){{\bf X}}
\put(192,178){{\bf Y}}
\put(280,150){SCALES}
\put(113,93){\dashbox{3}(155,155)}
\put(115,95){\dashbox{2}(70,70)}
\put(120,100){\dashbox{2}(20,20)}
\put(80,169){\tiny {${\it G}_3$}}
\put(90,169){\vector(1,-1){23}}
\put(80,156){\tiny {${\it G}_2$}}
\put(90,156){\vector(1,-1){25}}
\put(80,143){\tiny {${\it G}_1$}}
\put(90,143){\vector(1,-1){30}}
\put(80,130){\tiny {${\it G}_0$}}
\put(90,130){\vector(1,-1){30}}
\put(118,98){$\bullet$}
\put(120,95){\tiny {,0)}}
\put(138,118){$\bullet$}
\put(135,125){\tiny {,2)}}
\put(118,118){$\bullet$}
\put(120,125){\tiny {,3)}}
\put(138,98){$\bullet$}
\put(135,95){\tiny{,1)}}

\put(160,100){\dashbox{2}(20,20)}
\put(158,98){$\bullet$}
\put(160,95){\tiny {,1,0)}}
\put(178,118){$\bullet$}
\put(173,125){\tiny {,12)}}
\put(158,118){$\bullet$}
\put(158,125){\tiny {,1,3)}}
\put(178,98){$\bullet$}
\put(173,95){\tiny ,{1,1)}}

\put(195,95){\dashbox{2}(70,70)}
\put(200,100){\dashbox{2}(20,20)}
\put(198,98){$\bullet$}
\put(218,118){$\bullet$}
\put(198,118){$\bullet$}
\put(218,98){$\bullet$}

\put(240,100){\dashbox{2}(20,20)}
\put(238,98){$\bullet$}
\put(258,118){$\bullet$}
\put(238,118){$\bullet$}
\put(258,98){$\bullet$}

\put(160,140){\dashbox{2}(20,20)}
\put(118,138){$\bullet$}
\put(116,135){\tiny {,3,0)}}
\put(138,158){$\bullet$}
\put(133,165){\tiny {,3,2)}}
\put(118,158){$\bullet$}
\put(114,165){\tiny {,3,3)}}
\put(138,138){$\bullet$}
\put(135,135){\tiny ,{3,1)}}

\put(120,140){\dashbox{2}(20,20)}
\put(158,138){$\bullet$}
\put(160,135){\tiny {,2,0)}}
\put(178,158){$\bullet$}
\put(173,165){\tiny {,2,2)}}
\put(158,158){$\bullet$}
\put(158,165){\tiny {,2,3)}}
\put(178,138){$\bullet$}
\put(175,135){\tiny{,2,1)}}

\put(200,140){\dashbox{2}(20,20)}
\put(198,138){$\bullet$}
\put(218,158){$\bullet$}
\put(198,158){$\bullet$}
\put(218,138){$\bullet$}

\put(240,140){\dashbox{2}(20,20)}
\put(238,138){$\bullet$}
\put(258,158){$\bullet$}
\put(238,158){$\bullet$}
\put(258,138){$\bullet$}

\put(120,180){\dashbox{2}(20,20)}
\put(115,175){\dashbox{2}(70,70)}
\put(118,178){$\bullet$}
\put(138,198){$\bullet$}
\put(118,198){$\bullet$}
\put(138,178){$\bullet$}

\put(160,180){\dashbox{2}(20,20)}
\put(158,178){$\bullet$}
\put(178,198){$\bullet$}
\put(158,198){$\bullet$}
\put(178,178){$\bullet$}

\put(200,180){\dashbox{2}(20,20)}
\put(195,175){\dashbox{2}(70,70)}
\put(197,172){\tiny {$,1,0,0)$}}
\put(198,178){$\bullet$}
\put(218,198){$\bullet$}
\put(198,198){$\bullet$}
\put(218,178){$\bullet$}

\put(240,180){\dashbox{2}(20,20)}
\put(238,178){$\bullet$}
\put(258,198){$\bullet$}
\put(238,198){$\bullet$}
\put(258,178){$\bullet$}

\put(120,220){\dashbox{2}(20,20)}
\put(118,218){$\bullet$}
\put(138,238){$\bullet$}
\put(118,238){$\bullet$}
\put(138,218){$\bullet$}

\put(160,220){\dashbox{2}(20,20)}
\put(158,218){$\bullet$}
\put(178,238){$\bullet$}
\put(158,238){$\bullet$}
\put(178,218){$\bullet$}

\put(200,220){\dashbox{2}(20,20)}
\put(198,218){$\bullet$}
\put(218,238){$\bullet$}
\put(198,238){$\bullet$}
\put(218,218){$\bullet$}

\put(240,220){\dashbox{2}(20,20)}
\put(238,218){$\bullet$}
\put(258,238){$\bullet$}
\put(238,238){$\bullet$}
\put(258,218){$\bullet$}
\put(120,50){$|X-Y|_H$=$L^{scale(X,Y)}$=$L^3$}
\put(0,30){For $L=2$, $|X-Y|=(...,1,2,2)$=$2^3$; $X_i\in {\bf Z}_4$ and
${\it G}=\oplus^{\infty}_{k=0}{\bf Z}_4$.}
\put(160,10){1.b)}
\put(210,350){{\bf X}}
\put(240,350){{\bf Y}}
\put(120,320){$|X-Y|=L^{scale(X,Y)}$=$L^4$.}
\put(0,300){For $L=2$, $|X-Y|=(...,1,1,1,1)$=$2^4$;
$X_i\in {\bf Z}_2$ and ${\it G}=\oplus^{\infty}_{k=0}{\bf Z}_2$.}
\put(160,280){1.a)}
\put(0,360){$\bullet$}
\put(0,340){\tiny {$,0)$}}
\put(0,410){$\cdot$}
\put(0,420){\tiny{${\it G}_0$}}
\put(15,440){\tiny {${\it G}_1$}}
\put(0,430){\vector(1,0){30}}
\put(0,450){\vector(1,0){90}}
\put(45,460){\tiny {${\it G}_2$}}
\put(0,470){\vector(1,0){210}}
\put(105,480){\tiny {${\it G}_3$}}
\put(0,490){\vector(1,0){450}}
\put(225,500){\tiny {${\it G}_4$}}
\put(30,360){$\bullet$}
\put(30,340){\tiny {$,1)$}}
\put(60,360){$\bullet$}
\put(60,340){\tiny {$,1,0)$}}
\put(90,360){$\bullet$}
\put(90,340){\tiny {$,1,1)$}}
\put(120,360){$\bullet$}
\put(120,340){\tiny {$,1,0,0)$}}
\put(150,360){$\bullet$}
\put(150,340){\tiny {$,1,0,1)$}}
\put(180,360){$\bullet$}
\put(180,340){\tiny {$,1,1,0)$}}
\put(210,360){$\bullet$}
\put(210,340){\tiny {$,1,1,1)$}}
\put(240,360){$\bullet$}
\put(240,340){\tiny {$,1,0,0,0)$}}
\put(270,360){$\bullet$}
\put(300,360){$\bullet$}
\put(330,360){$\bullet$}
\put(360,360){$\bullet$}
\put(390,360){$\bullet$}
\put(420,360){$\bullet$}
\put(450,360){$\bullet$}
\put(0,370){- - - - - }
\put(60,370){- - - - - }
\put(120,370){- - - - - }
\put(180,370){- - - - - }
\put(240,370){- - - - - }
\put(300,370){- - - - - }
\put(360,370){- - - - - }
\put(420,370){- - - - - }
\put(0,380){- - - - - - - - - - - - - }
\put(120,380){- - - - - - - - - - - - - }
\put(240,380){- - - - - - - - - - - - -}
\put(360,380){- - - - - - - - - - - - - }
\put(0,390){- - - - - - - - - - - - - - - - - - - - - - - - - - - -}
\put(240,390){- - - - - - - - - - - - - - - - - - - - - - - - - - - - }
\put(0,400){- - - - - - - - - - - - - - - - - - - - - - - - - - - - -
- - - - - - - - - - - - - - - - - - - - - - - - - - - - -}
\put(400,430){ SCALES}
\end{picture}
\caption{}
\end{figure}

\newpage
\begin{figure}
\begin{picture}(500,500)(0,0)
\put(0,350){\framebox(20,20)}
\put(250,350){\dashbox{2}(20,20){$\bullet$}}
\put(40,350){\framebox(20,20)}
\put(290,350){\dashbox{2}(20,20){$\bullet$}}
\put(80,350){\framebox(20,20)}
\put(90,340){8}
\put(330,350){\dashbox{2}(20,20){$\bullet$}}
\put(120,350){\framebox(20,20)}
\put(130,340){12}
\put(370,350){\dashbox{2}(20,20){$\bullet$}}
\put(0,390){\framebox(20,20)}
\put(10,380){1}
\put(250,390){\dashbox{2}(20,20){$\bullet$}}
\put(40,390){\framebox(20,20)}
\put(50,380){4}
\put(290,390){\dashbox{2}(20,20){$\bullet$}}
\put(80,390){\framebox(20,20)}
\put(90,380){7}
\put(330,390){\dashbox{2}(20,20){$\bullet$}}
\put(120,390){\framebox(20,20)}
\put(130,380){11}
\put(370,390){\dashbox{2}(20,20){$\bullet$}}
\put(0,430){\framebox(20,20)}
\put(250,430){\dashbox{2}(20,20){$\bullet$}}
\put(40,430){\framebox(20,20)}
\put(50,420){3}
\put(290,430){\dashbox{2}(20,20){$\bullet$}}
\put(80,430){\framebox(20,20)}
\put(90,420){6}
\put(330,430){\dashbox{2}(20,20){$\bullet$}}
\put(120,430){\framebox(20,20)}
\put(130,420){10}
\put(370,430){\dashbox{2}(20,20){$\bullet$}}
\put(0,470){\framebox(20,20)}
\put(250,470){\dashbox{2}(20,20){$\bullet$}}
\put(40,470){\framebox(20,20)}
\put(50,460){2}
\put(290,470){\dashbox{2}(20,20){$\bullet$}}
\put(80,470){\framebox(20,20)}
\put(90,460){5}
\put(330,470){\dashbox{2}(20,20){$\bullet$}}
\put(120,470){\framebox(20,20)}
\put(130,460){9}
\put(370,470){\dashbox{2}(20,20){$\bullet$}}
\put(170,420){\vector(1,0){50}}
\put(170,325){\vector(-1,1){30}}
\put(170,320){${\it G}_1\in {\it G}$}
\put(270,320){$X'\in {\it G}'\sim {\it G}$}
\put(360,328){\vector(-1,1){20}}
\put(360,320){contracted ${\it G}_1$}
\put(190,300){2.a)}
\put(284,330){\vector(1,2){13}}
\put(170,430){$R(w)=w'$}
\put(0,210){\framebox(20,20)}
\put(30,220){=}
\put (60,270){\framebox(20,20)}
\put(90,280){$O(\lambda)$}
\put(60,230){\framebox(20,20)}
\put(90,240){$O(\lambda^2)$}
\put(60,190){\framebox(20,20)}
\put(90,200){$O(\lambda^3)$}
\put(70,180){$\cdot$}
\put(70,170){$\cdot$}
\put(70,160){$\cdot$}
\put(70,120){2.b)}
\put(160,210){\framebox(20,20)}
\put(190,220){=}
\put(220,270){\framebox(20,20)}
\put(250,280){$O(\lambda)$}
\put(220,230){\framebox(20,20)}
\put(250,240){$O(\lambda^2)$}
\put(220,190){\framebox(20,20)}
\put(250,200){$O(\lambda^3)$}
\put(230,180){$\cdot$}
\put(230,170){$\cdot$}
\put(230,160){$\cdot$}
\put(230,120){2.c)}
\put(320,210){\framebox(20,20)}
\put(350,220){=}
\put(380,270){\framebox(20,20)}
\put(410,280){$O(\lambda^2)$}
\put(380,230){\framebox(20,20)}
\put(410,240){$O(\lambda^3)$}
\put(380,190){\framebox(20,20)}
\put(410,200){$O(\lambda^4)$}
\put(390,180){$\cdot$}
\put(390,170){$\cdot$}
\put(390,160){$\cdot$}
\put(390,120){2.d)}
\put(0,50){${\it G}_1\in {\it G}$ marked in}
\put(0,30){ Figure 2.a) like}
\put(150,60){1,2,,4,5,8,9,10,11,12}
\put(150,40){3,7}
\put(150,20){6}
\put(260,40){are}
\put(300,60){Figure 2.b) type}
\put(300,40){Figure 2.c) type}
\put(300,20){Figure 2.d) type}
\end{picture}
\caption{}
\end{figure}
\end{document}